\documentclass[a4paper,11pt]{article}
\pdfoutput=1 

\usepackage{jinstpub} 

\usepackage[T1]{fontenc} 

\usepackage{graphicx}
\usepackage{color}
\usepackage{mathtools}
\usepackage[left]{lineno}
\usepackage{hyperref}
\usepackage{subfigure}
\usepackage{gensymb}
\usepackage{wrapfig}

\usepackage{natbib}

\title{R2D2 spherical TPC: first energy resolution results}

\author[a]{R.~Bouet}
\author[b]{J.~Busto}
\author[a,f]{V.~Cecchini}
\author[a]{C.~Cerna}
\author[c]{A.~Dastgheibi-Fard}
\author[a]{F.~Druillole}
\author[a]{C.~Jollet}
\author[a]{P.~Hellmuth}
\author[d]{I.~Katsioulas}
\author[d,e]{P.~Knights}
\author[e]{I.~Giomataris}
\author[e]{M.~Gros}
\author[f]{P.~Lautridou}
\author[a,1]{A.~Meregaglia\note{Corresponding author}}
\author[e] {X.~F.~Navick}
\author[d]{T.~Neep}
\author[d]{K.~Nikolopoulos}
\author[a]{F.~Perrot}
\author[a]{F.~Piquemal}
\author[a]{M.~Roche}
\author[a]{B.~Thomas}
\author[d]{R.~Ward}
\author[c]{M.~Zampaolo}

\affiliation[a]{CENBG, Universit\'{e} de Bordeaux, CNRS/IN2P3, F-33175 Gradignan, France}
\affiliation[b]{CPPM, Universit\'{e} d'Aix-Marseille, CNRS/IN2P3, F-13288 Marseille, France}
\affiliation[c]{LSM, CNRS/IN2P3, Universit\'{e} Grenoble-Alpes, Modane, France}
\affiliation[d]{School of Physics and Astronomy, University of Birmingham, B15 2TT, United Kingdom}
\affiliation[e]{IRFU, CEA, Universit\'{e} Paris-Saclay, F-91191 Gif-sur-Yvette, France}
\affiliation[f]{SUBATECH, IMT-Atlantique, Universit\'{e} de Nantes, CNRS-IN2P3, France}

\abstract{Spherical time projection chambers (TPC), also known as
spherical proportional counters, are employed in the search for rare phenomena,
such as light Dark Matter candidates. The spherical TPC exhibits a number of essential features, making it a promising candidate for the search of neutrinoless double
beta decay ($\beta\beta0\nu$). A tonne-scale spherical TPC experiment could cover a region of parameter space relevant for the
inverted mass hierarchy with a few years of data taking. In this direction, the major R\&D goal of the R2D2 effort is the demonstration of the required  energy resolution. First results from an argon-filled prototype detector are reported, demonstrating an energy
resolution of 1.1\% FWHM for 5.3~MeV $\alpha$ tracks in the 0.2 to 1.1~bar pressure range. 
This is a major milestone in terms of energy resolution, paving the way for further studies with
xenon gas, and the possible use of this technology for $\beta\beta0\nu$
searches.}

\begin{document}
\maketitle
\flushbottom

\section{Introduction}

The Spherical Proportional Counter (SPC), or spherical time projection chamber (TPC), is a novel gaseous detector concept which combines several promising features including good energy resolution, low energy threshold, low background capability, and the possibility of instrumenting large target masses. The detector was invented and initially developed in CEA Saclay~\cite{Giomataris:2008ap}, primarily aiming to study low energy neutrino physics: neutrino oscillations, neutrino coherent elastic scattering and Supernova neutrino detection~\cite{Giomataris:2003bp,Giomataris:2005fx}.

The detector consists of a large grounded spherical metallic vessel, which acts as the cathode, 
and a small metallic spherical anode, with  a diameter of order 1~mm at the centre. 
This simple and robust structure allows large volumes to be readout with a single electronic channel. Thanks to its low capacitance and high gas gain, the detector is capable of measuring very low energy depositions, down to single electrons. Large target masses can be achieved, by operating large size detectors at high gas pressure.

SPCs are employed in the search for light Dark Matter candidates, in the mass range between 0.1 and 10 GeV, by the NEWS-G
collaboration~\cite{Gerbier:2014jwa,Arnaud:2017bjh,Savvidis:2016wei}. The possibility to achieve
a per cent level  energy resolution, the simplicity of the
detector with only one readout channel, and the low material budget, make SPCs an appealing option for the search of  other rare phenomena, such as
neutrinoless double beta decays ($\beta\beta0\nu$). This possibility
has been investigated for a $^{136}$Xe-filled detector at a
pressure of 40~bars~\cite{Meregaglia:2017nhx}, suggesting sensitivity to the inverted mass hierarchy region, leading to the establishment of the Rare Decays with Radial Detector (R2D2) R\&D effort.

The first aim of the R\&D effort is the demonstration of an
energy resolution of 1\% FWHM at 2.458~MeV, which corresponds to the transition
energy ($Q_{\beta\beta}$) of the $^{136}$Xe double beta decay. To validate this critical aspect, and to fully characterise
the detector response, a prototype has been
designed and constructed at CENBG: it has been operated with a gas mixture of 98\% argon and 2\%
methane.

The main goal of this paper is to present the first energy resolution measurements obtained with the SPC prototype in argon: a value at the level of 1.1\% was obtained with 5.3~MeV $\alpha$ tracks. This is the first time such an excellent energy resolution is achieved for an SPC in the MeV energy range. It is also shown that the resolution does not depend on the track length, a critical feature for the $\beta\beta0\nu$ decay search. 
Furthermore, detailed simulations are compared to the experimental results to understand the detector response.
Signal processing
techniques are also explored, aiming to exploit the detector characteristics and maximise the obtained information, for example to infer the track direction and position. 


\section{Experimental setup}
\subsection{Detector description}
\begin{figure} [t]
\centering
\subfigure[\label{fig:det1}]{\includegraphics[height=8
cm]{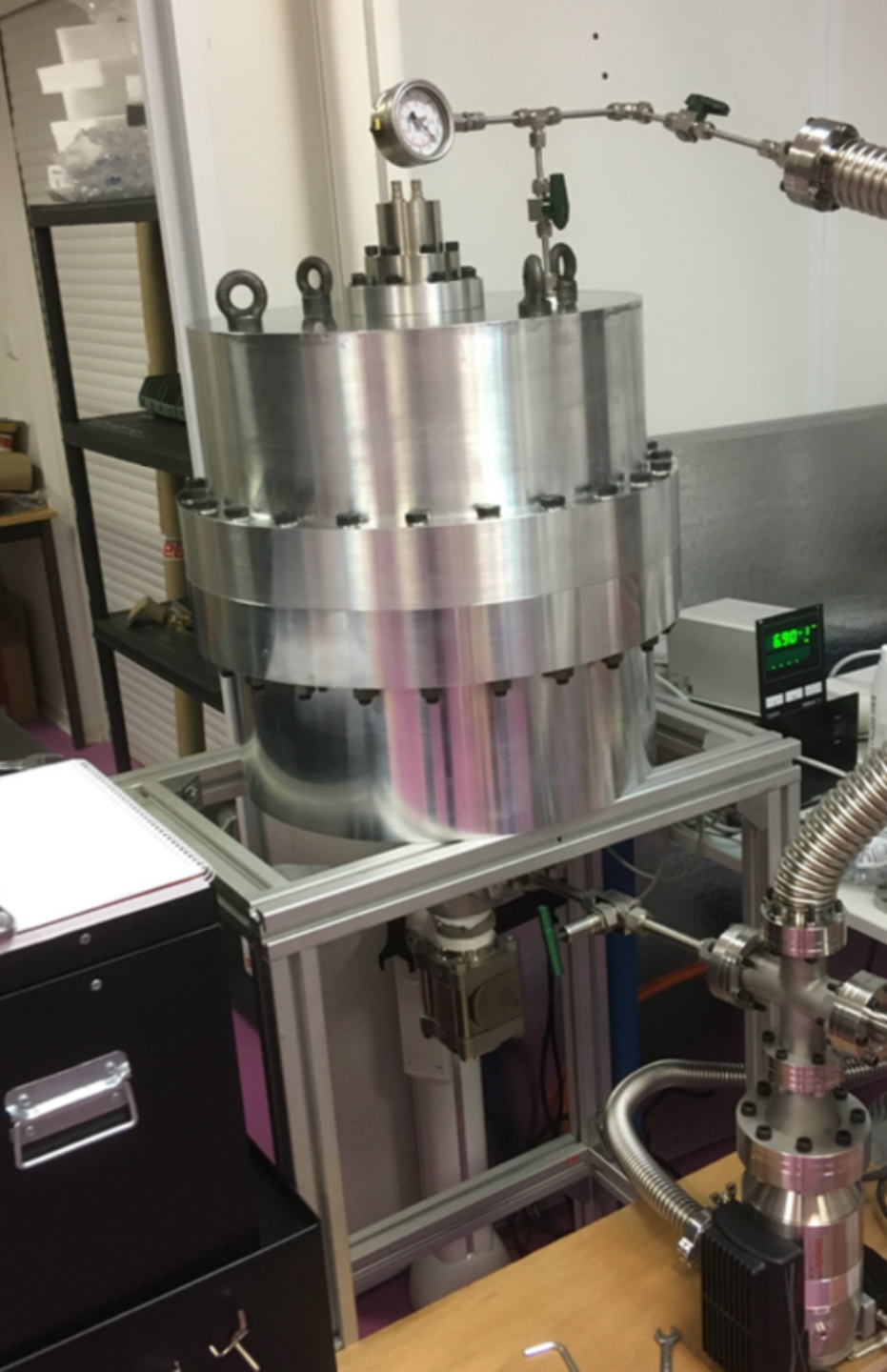}}
\subfigure[\label{fig:det2}]{\includegraphics[height=8
cm]{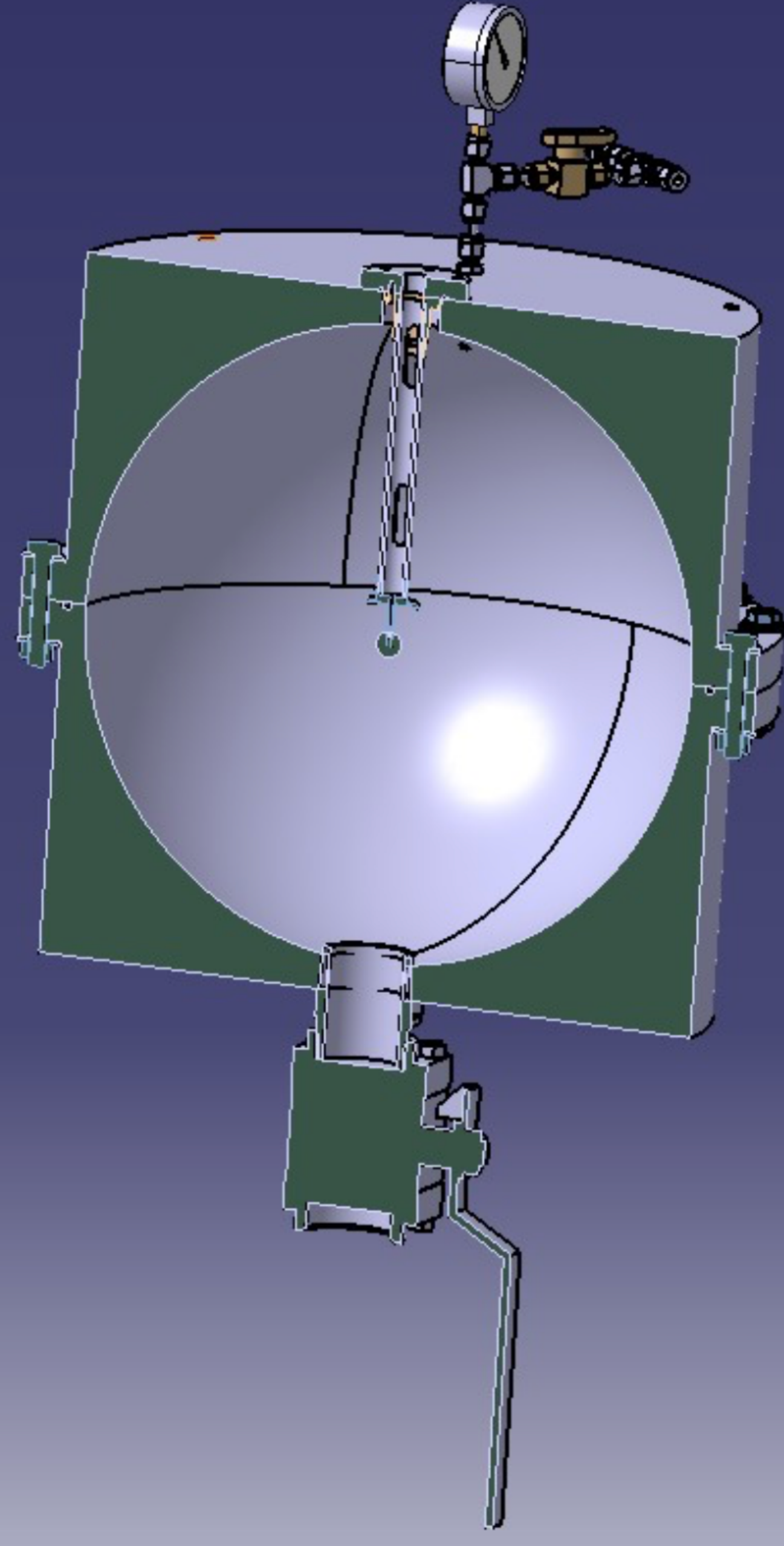}} \caption{{\it \subref{fig:det1}~Actual realisation
and \subref{fig:det2}~mechanical drawing of the R2D2 detector. The two
hemispheres can be seen as well as the pressure gauge, the HV connectors on top
of the detector, and the large valve at the bottom to allow for the insertion of
a radioactive source. }}
\label{fig:1}
\end{figure}

The R2D2 setup consists of a 20~cm radius sphere 
which could contain about 10~kg of xenon at a pressure of 40~bar. Given
that high-radiopurity is not critical at this R\&D stage,
aluminium was chosen for the detector construction. This choice was driven by
the in-house capabilities of CENBG.
The detector was built starting from two cylindrical blocks of
aluminium which were machined in order to form two hemispheres, and then
bolted together as shown in Fig~\ref{fig:1}.

The anode, having a positive high voltage, is located at the centre of the grounded sphere, leading to the drift of the ionisation electrons and their subsequent
amplification to form the observed signal. The sensor~\cite{Katsioulas:2018pyh} consists of a 2~mm diameter stainless-steel sphere
connected to an insulated wire of 150~$\mu$m diameter, which is held in place
by a grounded supporting rod. 

Readout electronics play a critical role, given the stringent requirements in terms of low noise and immunity to radio frequency
 interferences (RFI) in order to achieve a good energy resolution. Details on the specific developments are given in
 Sec.~\ref{sec:elec}, whereas the main elements of the electronic chain are
 listed below:
 \begin{itemize}
\item {\it High-voltage splitter}. The high-voltage (HV) cable connecting the sensor to the power supply is also used to extract the signal. To separate the signal a dedicated splitter box has been prepared at CENBG.
\item {\it Preamplifier}. A preamplifier for impedance matching, shaping, and
amplification of the signal is required before passing it to the acquisition card.
A custom-made low-noise resistive-feedback charge-sensitive
preamplifier is used, with an adapted-frequency bandwidth, designed at CENBG as part of the OWEN (Optimal Waveform recognition Electronic Node) project~\cite{OWEN}.
\item {\it High-voltage power supply}. HV power supply can be an important 
source of noise. A commercial CAEN power supply, DT8034~\cite{CAEN}, is used, with a current monitor at 0.5~nA level and a voltage ripple below 10~mV.
\item {\it Data acquisition}. The data acquisition (DAQ) is done using a CALI card
read by the SAMBA acquisition software~\cite{Armengaud:2017rzu}.
\end{itemize}

\begin{figure}[t]
\centering
\subfigure[\label{fig:ele11}]{\includegraphics[height=6cm]{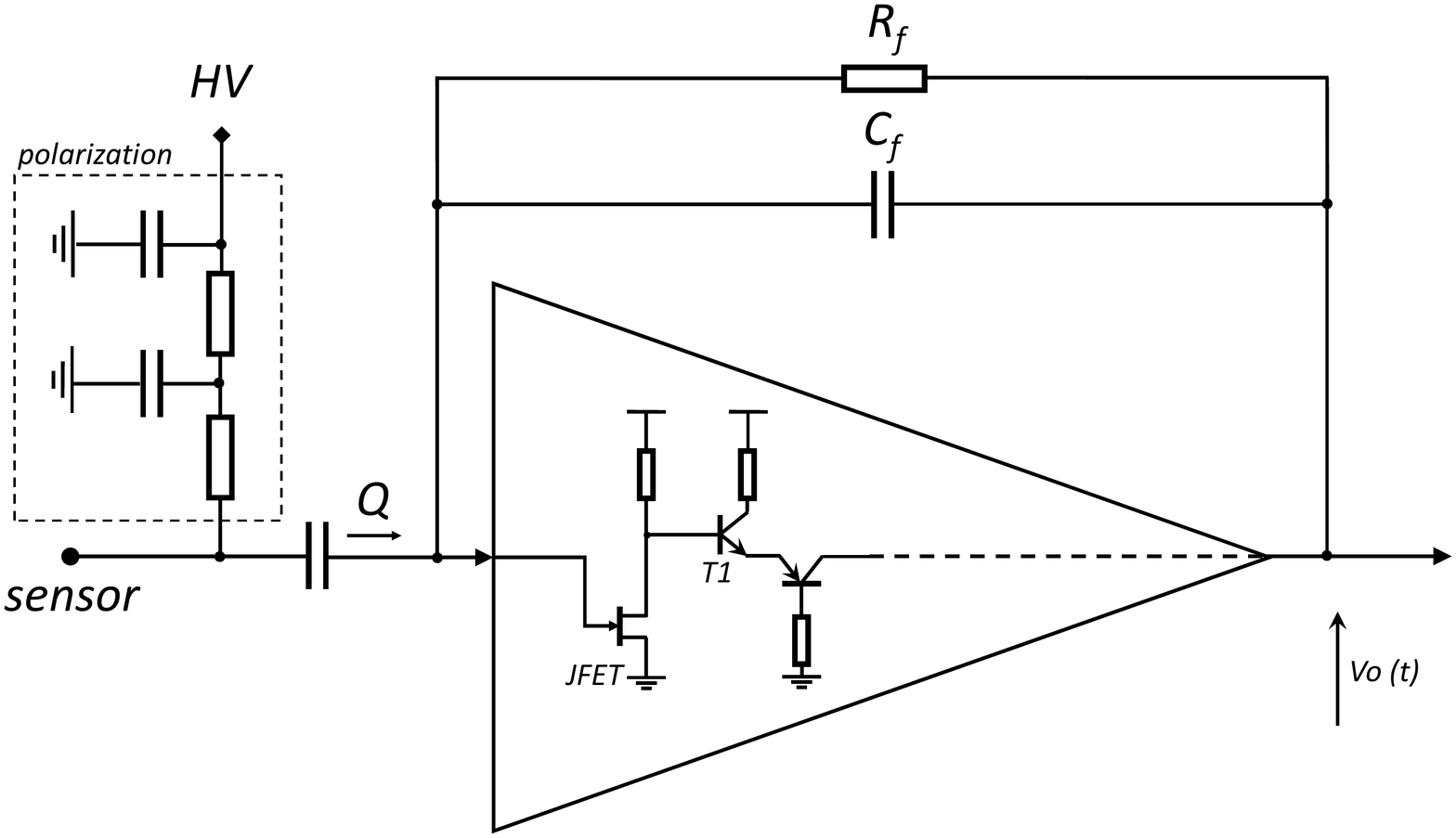}}
\subfigure[\label{fig:ele12}]{\includegraphics[height=6cm]{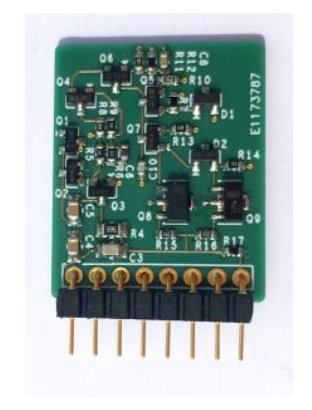}}
\caption{{\it \subref{fig:ele11} Simplified circuit of the OWEN charge sensitive
preamplifier with polarisation. The dashed box represents the HV splitter box.
\subref{fig:ele12}  Final realisation of the OWEN charge sensitive
preamplifier.\label{fig:ele1}}}
\end{figure}

\subsection{Electronics}
\label{sec:elec}

\begin{table}[tp]
\centering
\begin{tabular}{|c|c|}
\hline
Gain &  0.75 V/pC\\
Customizable gain & Yes\\
Feedback $C_f$ &  1.5 pF \\
Feedback $R_f$ &  100 M$\Omega$ \\
Decay time &  150 $\mu$s\\
Baseline noise &  920 $\mu$V pp\\
Risetime &  93 ns\\
SNR &  1868\\
Linearity &  1 fC to 2 pC\\

\hline
\end{tabular}
\caption{{\it Full specification of the OWEN preamplifier.}}

\label{tab:ele1}
\end{table}%

A charge-sensitive preamplifier was developed at CENBG in the framework of the
OWEN project~\cite{OWEN}. This circuit integrates the current signal from the sensor on the feedback
capacitance $C_f$, then generates an output-voltage signal which is
proportional to the original input charge $Q$ registered by the central anode. 
%
%
 The architecture of the preamplifier is based on the established design by
 T.V.~Blalock~\cite{ele}. Usually, the preamplifier uses a Field Effect
 Transistor (FET) as an input element associated with a bipolar transistor to
 form a cascode. In order to improve the gain bandwidth product (GBW), a
 transistor {\it T1} used as a current amplifier is inserted between the FET and the
 bipolar transistors, as shown in Fig.~\ref{fig:ele11}.

The layout of the preamplifier shown in Fig.~\ref{fig:ele12} has been optimized
in terms of noise, frequency bandwidth, and parasitic capacitance. The pinout is
pin compatible with the CREMAT \mbox{CR-110-R2}~\cite{CREMAT}, which was used for the
first tests of detector stability. The
preamplifier is coupled to a high-voltage filter shown in the dashed box of
Fig.~\ref{fig:ele11}.

The specifications of the OWEN preamplifier can be found in
Tab.~\ref{tab:ele1}. The noise performance was characterised with dedicated measurements. The HV splitter was based on an existing one developed for
the SEDINE~\cite{Fard:2015pla} detector and an optimization for the R2D2
detector is foreseen for a possible further reduction of the electronic noise.
The gain of the CREMAT preamplifier is 1.4~V/pC whereas for the OWEN
preamplifier a gain of 0.75~V/pC was measured. This difference is due to an
additional amplification stage which was avoided in the OWEN preamplifier. The
OWEN preamplifier meets the low noise requirements needed by the R2D2 detector
and it was therefore chosen as the baseline option.  In addition, the OWEN
preamplifier has the advantage of having customizable parameters allowing for a
better matching between the filter and the preamplifier bandwidth acceptance.

\subsection{Operation}
\label{sec:operation}
\begin{figure} [t]
\centering
   \subfigure[\label{fig:1bis}]{\includegraphics[height=7cm]{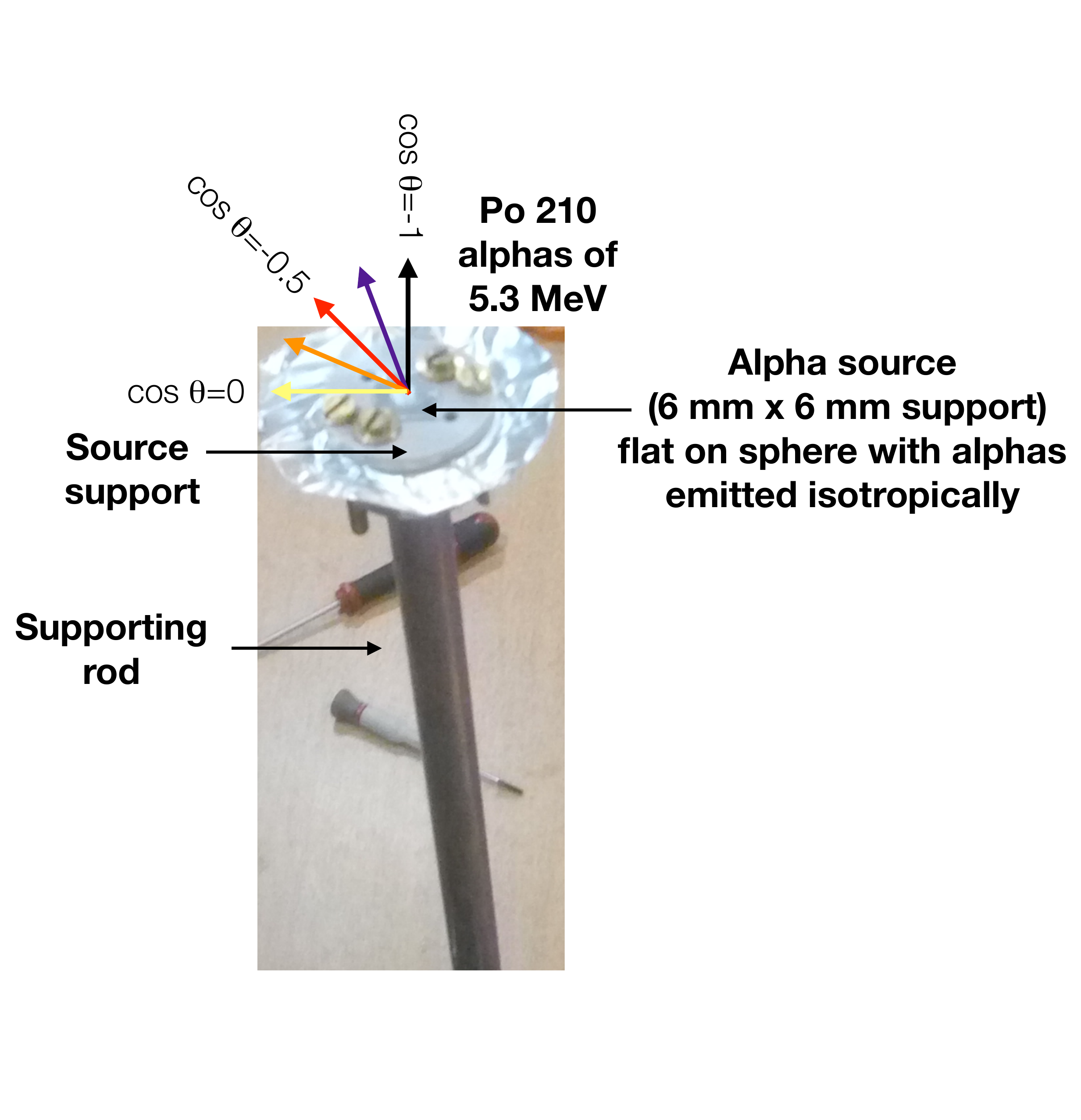}}
    \subfigure[\label{fig:1tris}]{\includegraphics[height=7cm]{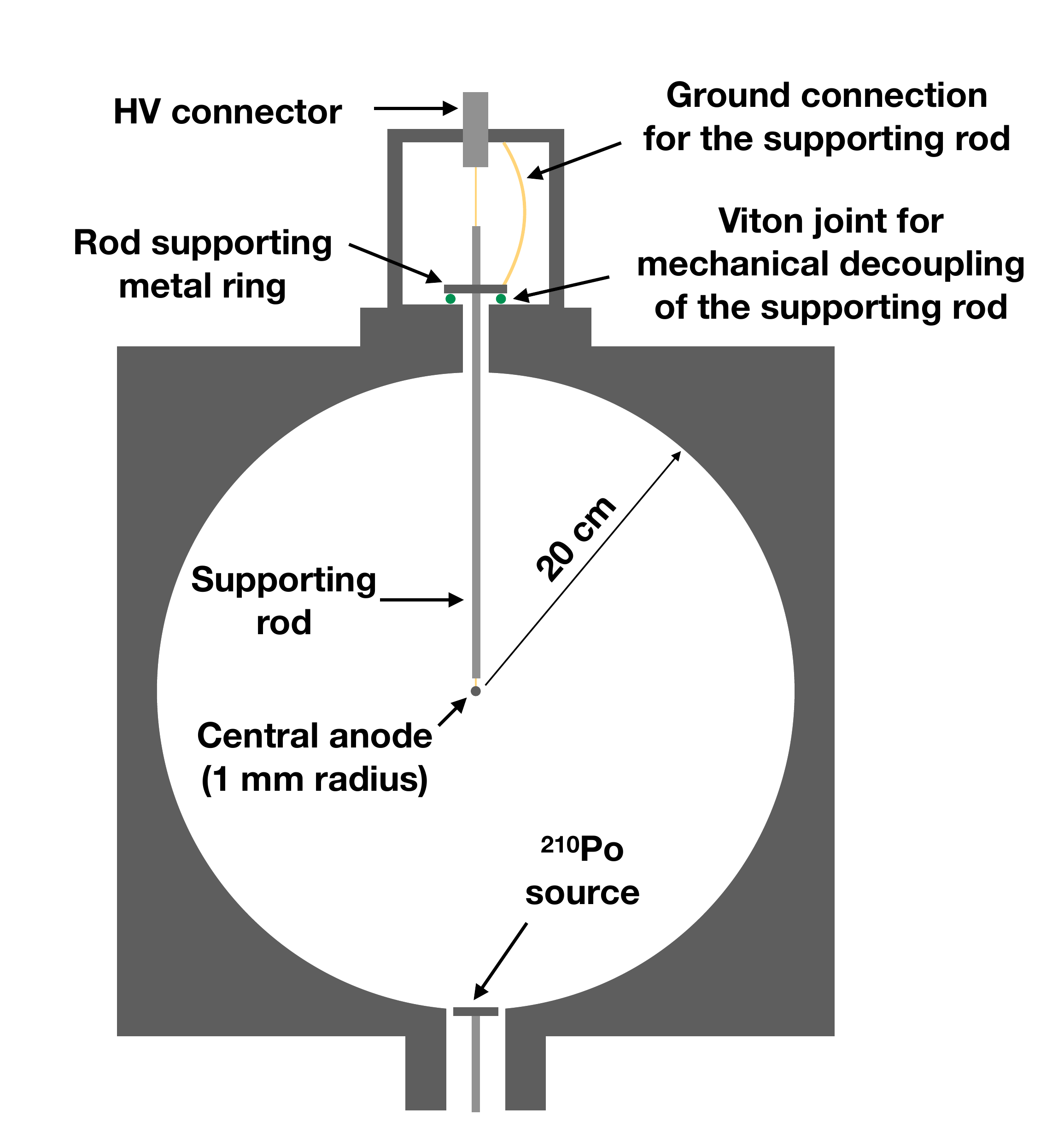}}
    \caption{{\it  \subref{fig:1bis} $^{210}$Po $\alpha$ source setup. The
    colour code represents the initial direction of the emitted
    $\alpha$-particles which corresponds to the one used in the plots showing
    the output of the simulation. \subref{fig:1tris} Schematic drawing of the
    experimental setup showing the central sensor support and the $^{210}$Po
    $\alpha$ source.}}
\end{figure}

For a good detector performance, contamination of the gas volume with
electronegative impurities has to be minimised. In particular the presence of
oxygen would result in a loss of signal which would be more important for electrons drifting longer distances, thus degrading the energy resolution. The used gas mixture 
is certified to have 
contamination at the level of 1~ppm, which is expected to be sufficient for these studies.

The detector has been heated for several weeks at 80\degree C in order to reduce the material outgassing during operation, and it is systematically pumped to a vacuum at the level of 10$^{-6}$~mbar before each filling. 
A vacuum of 10$^{-6}$~mbar corresponds to a purity of 5 ppb at 1 bar, which is negligible with respect to the gas purity of 1 ppm. 
The vacuum tightness was measured with an helium-leak tester and no leak was observed at the level of $5\times10^{-9}$~mbar/s. The main contribution to vacuum loss is due to outgassing of materials and a value of $2 \times10^{-6}$~mbar/s was measured, which ensures 
good detector operation for several weeks.
To test the detector response and resolution, a radioactive $\alpha$ source was used. Although the final experiment will observe electrons, it is  challenging to contain them in the current detector volume at low pressures, therefore, $\alpha$ particles were used, despite their ionisation
quenching~\cite{lindhard1963integral}.

A $^{210}$Po source producing $\alpha$ particles of 5.3~MeV with an activity of 4~Bq was used, allowing to quickly evaluate the detector gas gain. The main drawback of such a source is the electric field distortion due to the source itself and its support, shown in Fig.~\ref{fig:1bis}). 
Although the source is on the outer surface (see Fig.~\ref{fig:1tris}) to minimise potential perturbation of the electric
field, a weak distortion could still be present being more relevant for short alpha particle tracks i.e. at high pressure. Furthermore some $\alpha$ particles lose part of the energy in the source casing yielding a low energy tail which may affect the resolution.

The detector is sensitive to acoustic and electronic noise, as well as to
temperature variations. To work in the best available conditions, the R2D2
prototype was installed at the PRISNA facility~\cite{PRISNA} at CENBG, where the
temperature is kept constant within $1 ^\circ\mathrm{C}$ and human activity is reduced as much as possible. All the electronic devices were grounded to a large metallic plate on
which the detector is hosted.
To reduce vibrational noise impacting the baseline
stability, and therefore the resolution, the sensor supporting rod has been
mechanically decoupled from the rest of the detector through a joint as shown in
Fig.~\ref{fig:1tris}. Due to such a mechanical decoupling, a sizable impact on the noise reduction at low frequencies
was observed: a factor of about 2 at 60~Hz. For the next prototype, a
dedicated system to reduce vibration on the central sensor will be developed.

\section{Simulation}
\label{sec:Simulation}

\begin{figure}[t]
\centering
\subfigure[\label{fig:vdrift}]{\includegraphics[width=0.32\textwidth]{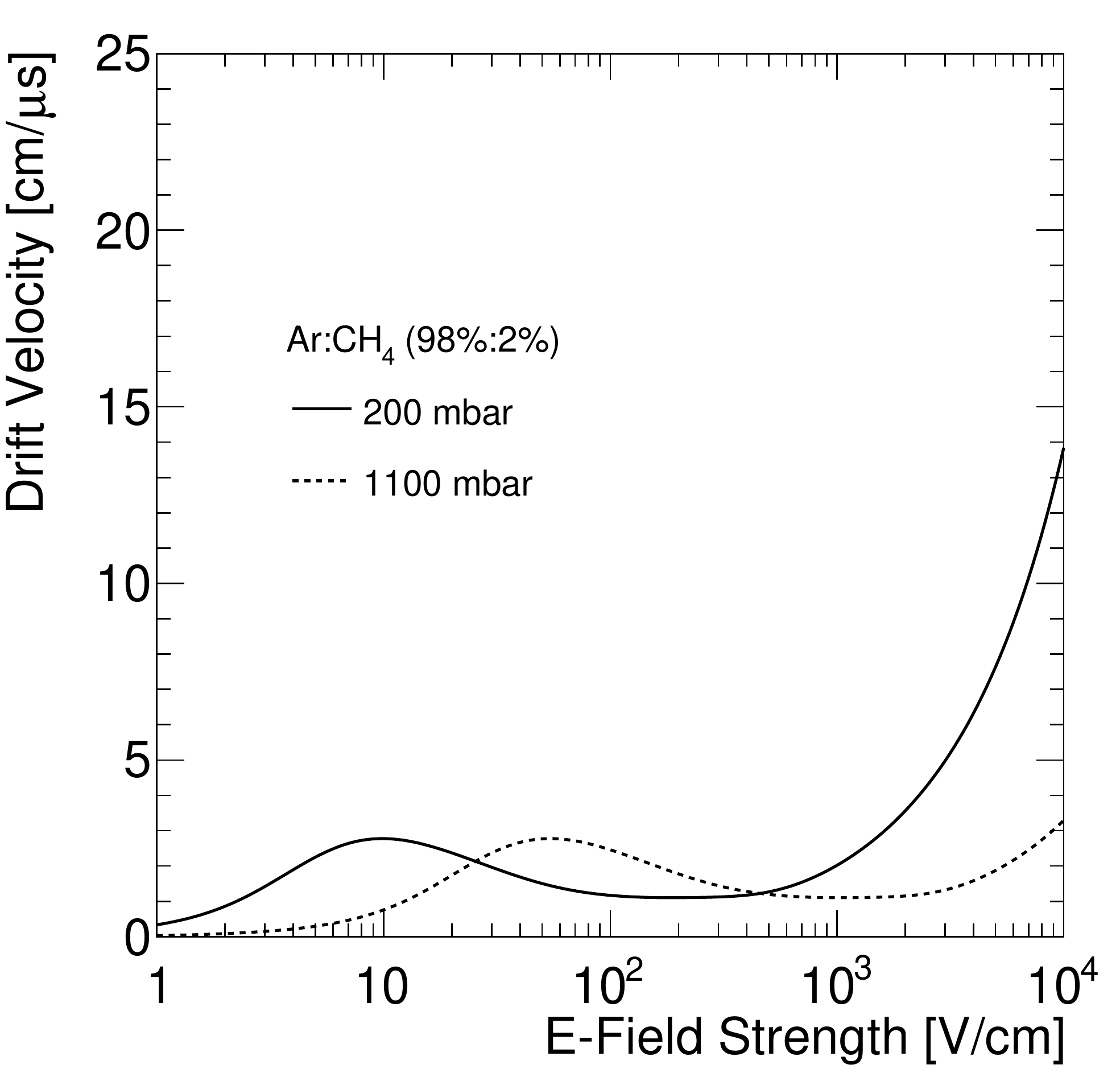}}
\subfigure[\label{fig:transDiff}]{\includegraphics[width=0.32\textwidth]{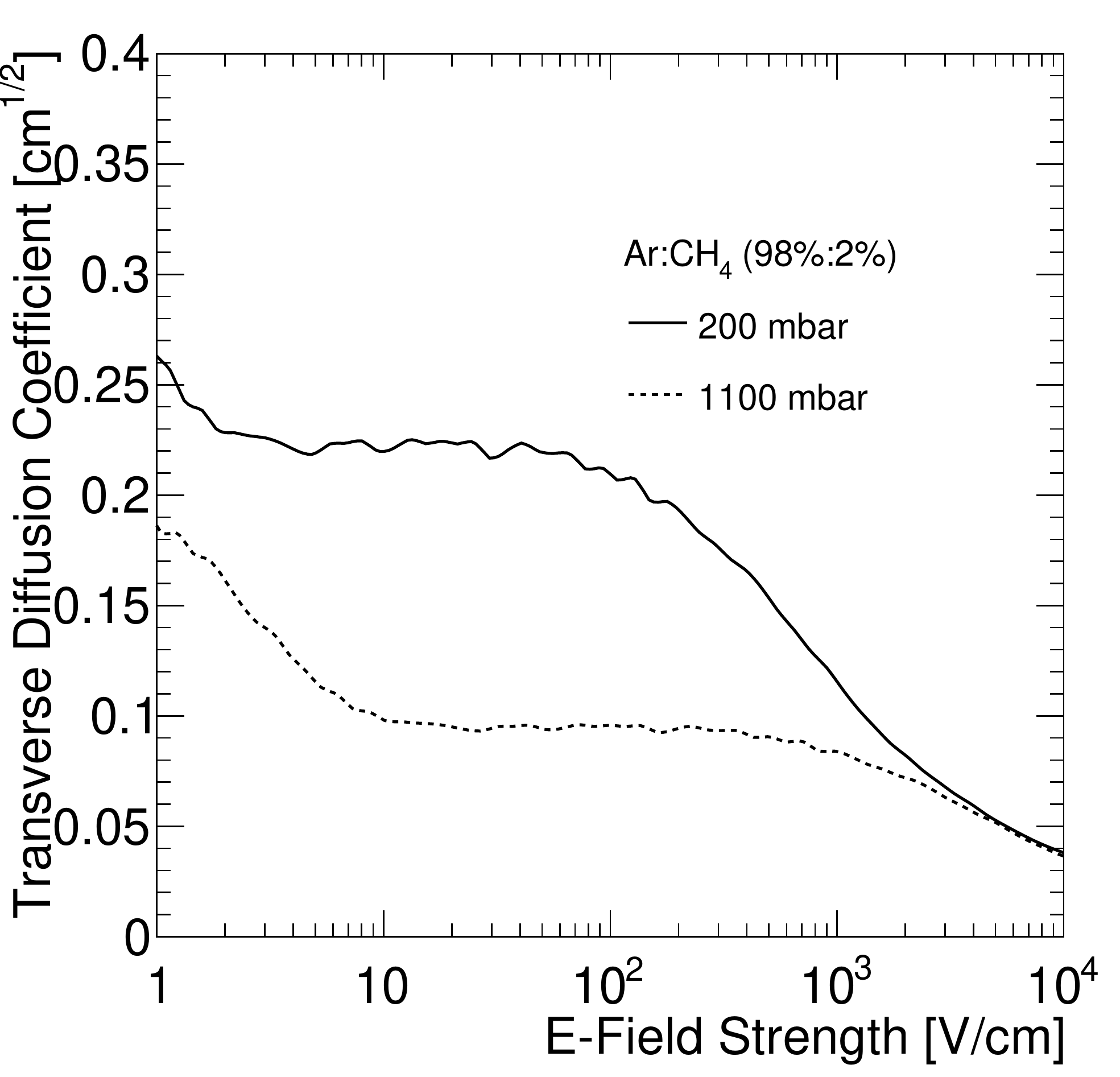}}
\subfigure[\label{fig:longDiff}]{\includegraphics[width=0.32\textwidth]{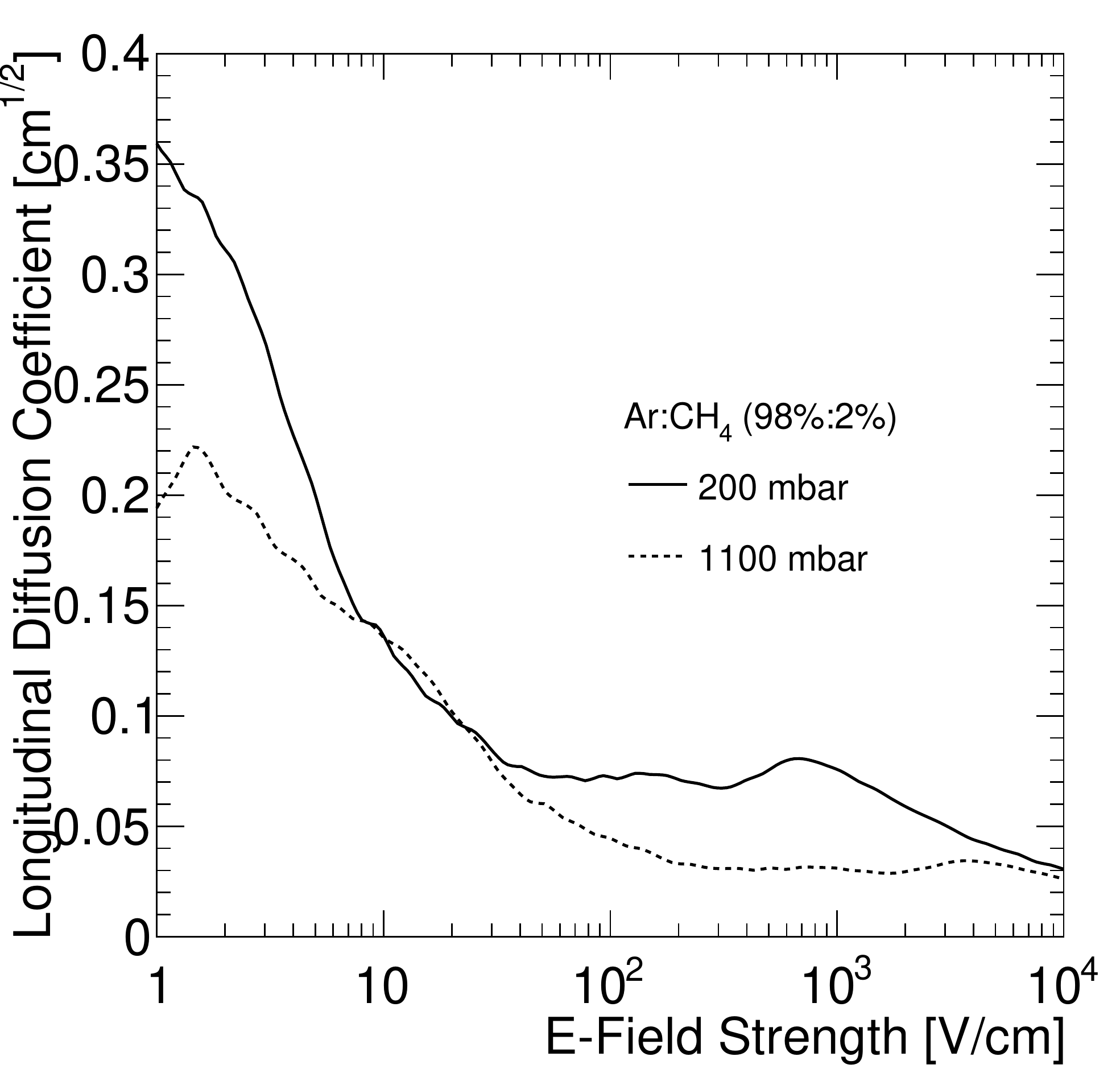}}
\caption{{\it Electron transport parameters (\subref{fig:vdrift} drift velocity, \subref{fig:transDiff} transverse and \subref{fig:longDiff} longitudinal diffusion coefficients) as a function of the electric field in Ar:CH$_4$ 98\%:2\% as estimated by Magboltz at 200~mbar and 1.1~bar.}}
\label{fig:transportParameters}
\end{figure}

\begin{figure}[t]
\centering
\includegraphics[width=0.55\textwidth]{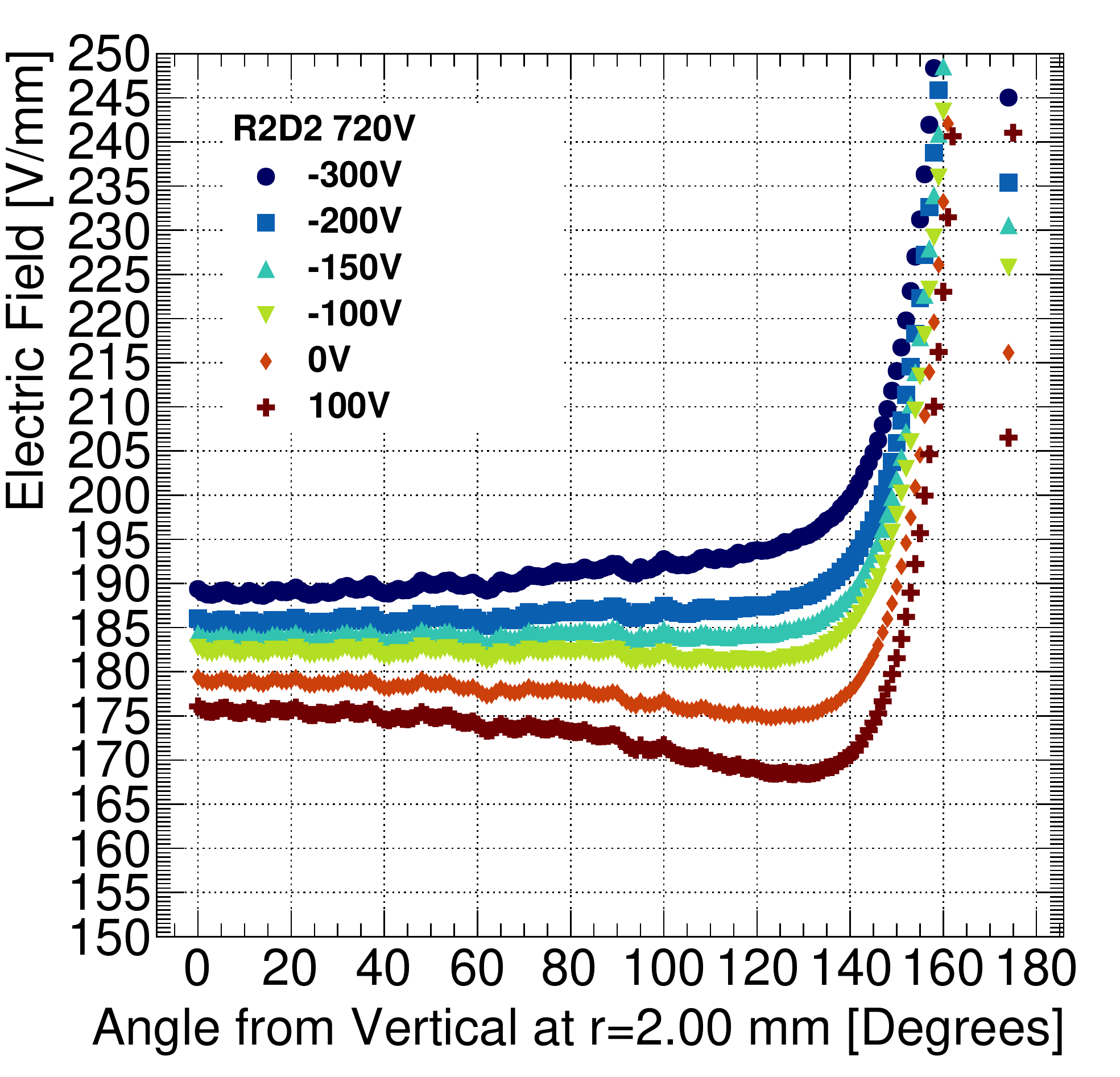} 
\caption{{\it Electric field strength $r = 2$ mm versus angle from the vertical.}\label{fig:correction_voltage:a}}
\end{figure}

The framework presented in Ref.~\cite{Katsioulas:2020ycw} has been used to
simulate the experimental set-up, combining Geant4~\cite{Allison:2016lfl}, a toolkit for the simulation
of particle interactions with matter, and Garfield++~\cite{Veenhof:1998tt, garfield}, a
toolkit for the detailed simulation of gaseous particle
detectors. Garfield++ interfaces to Heed~\cite{heed}, for
particle interactions, and Magboltz~\cite{magboltz}, for modelling electron
transport parameters in gases. The electric field in the
detector is described using the ANSYS~\cite{ansys} finite element software.


In Fig.~\ref{fig:transportParameters}, the drift velocity, as well as the
longitudinal and transverse diffusion coefficients are presented for the two gas
pressures considered in this study i.e. 200~mbar and 1.1~bar.
The gain of the gas mixture is typically governed by the Townsend coefficient, however, in gas mixtures like Ar:CH$_4$ the Penning effect is observed. The
methane ionisation potential is lower than the ionisation potential of argon,
and de-excitations of the latter may lead to ionisation of the former. In
the calculations a transfer probability of 15\%~\cite{Sahin:2010ssz} was used.

Typically, the sensor design includes a correction electrode. A voltage is applied to this electrode aiming to improve the electric field uniformity in the hemisphere near the anode support rod. This is optimised by studying
the electric field strength as a function of the zenith angle, 
at a fixed distance from the centre of the detector. This is presented in Figure~\ref{fig:correction_voltage:a}, where a clear variation 
in the electric field strength versus 
angle when no correction voltage is applied can be seen, and an optimal correction of approximately $-150$~V is obtained. 

\begin{figure}[t]
\centering
\subfigure[\label{fig:attachment1}]{\includegraphics[width=0.45\textwidth]{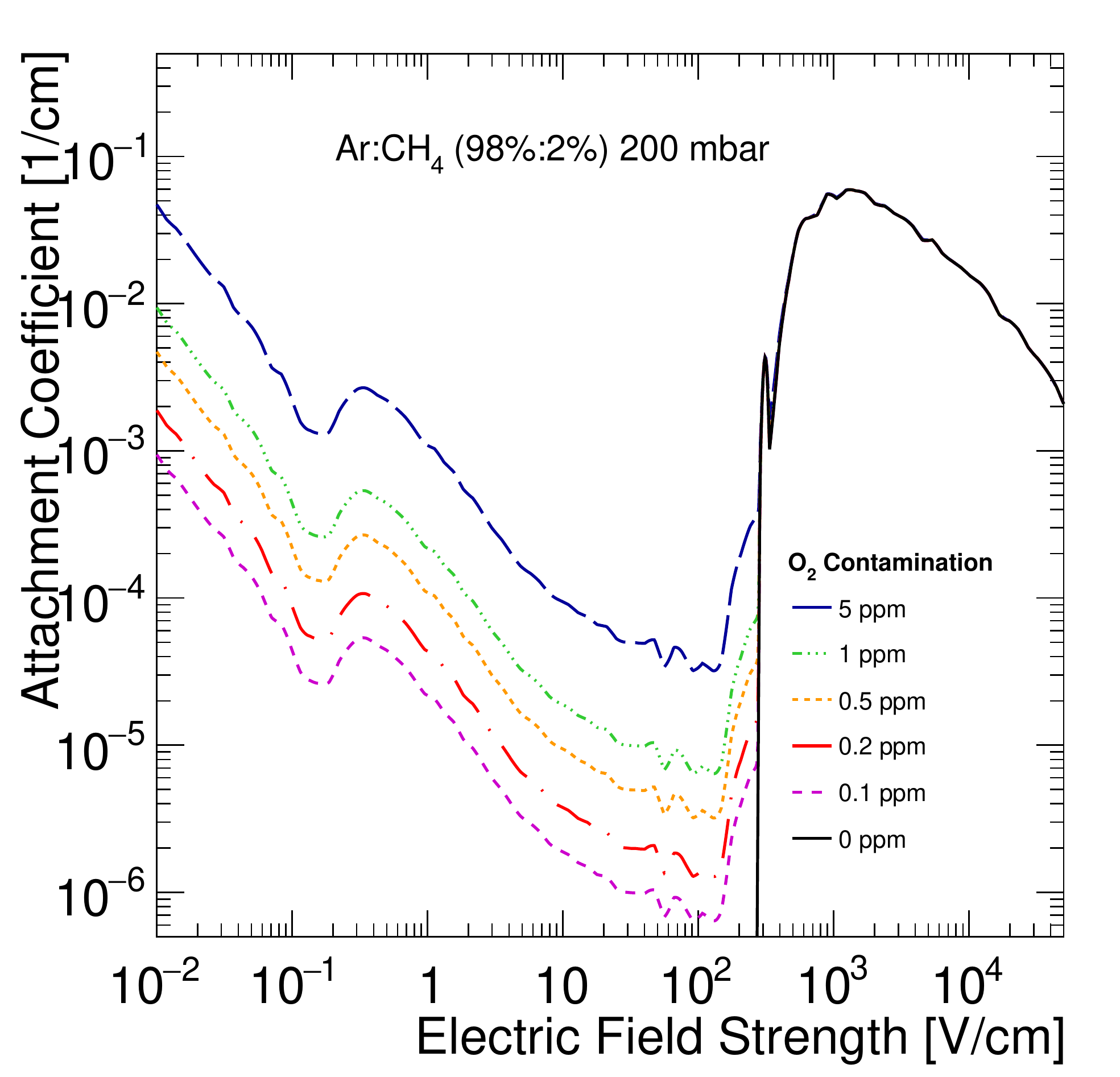}}
\subfigure[\label{fig:attachment2}]{\includegraphics[width=0.45\textwidth]{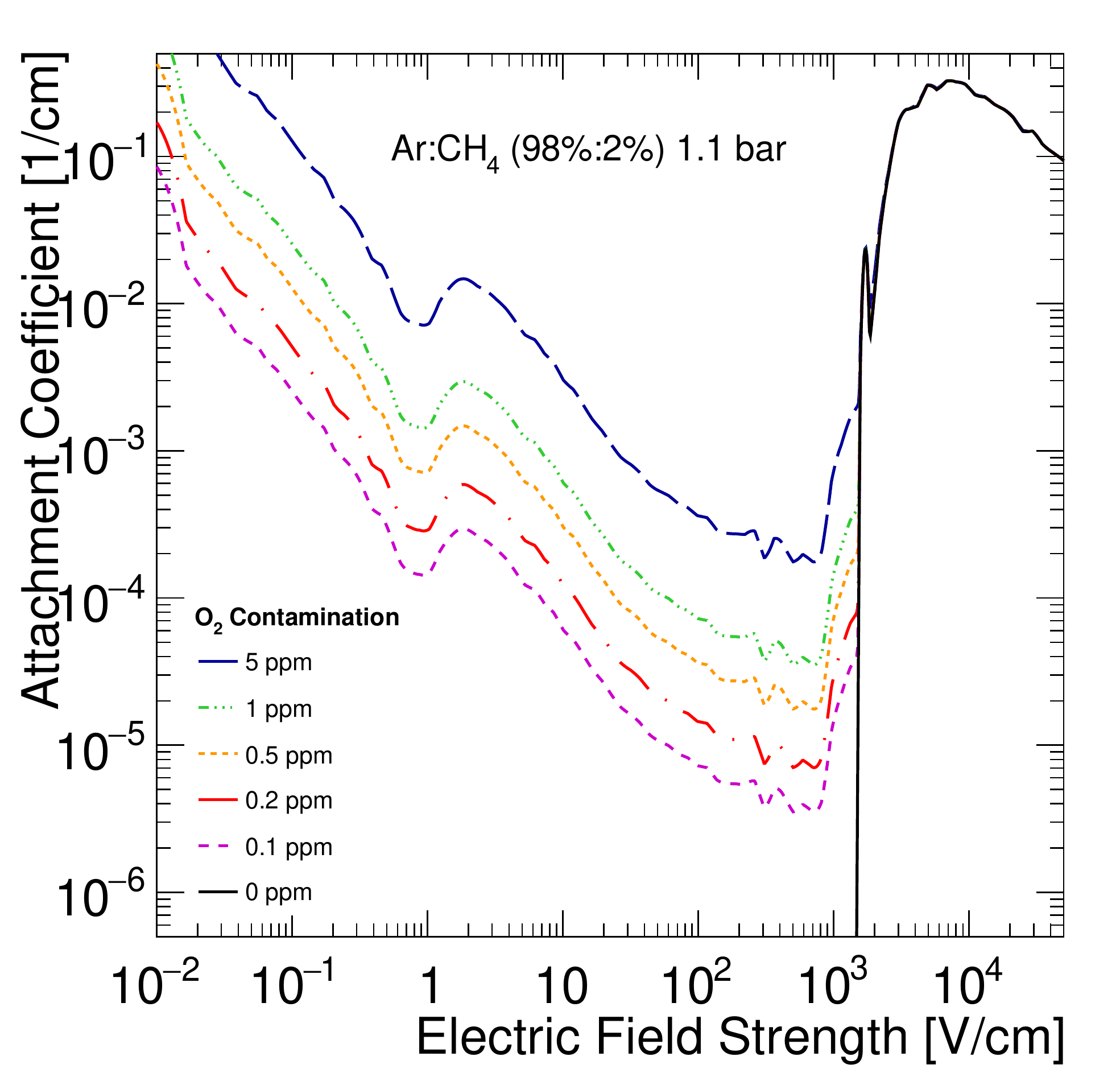}}
\caption{{\it Attachment coefficient as a function of electric field strength for different levels of $O_2$ contamination. \subref{fig:attachment1} for Ar:CH$_4$ 98\%:2\% at 200~mbar, and
\subref{fig:attachment2} for the same gas mixture at a pressure of 1.1~bar.}\label{fig:attachment}}
\end{figure}

\begin{figure}[h!]
\centering
\subfigure[\label{fig:impurity1}]{\includegraphics[width=0.45\textwidth]{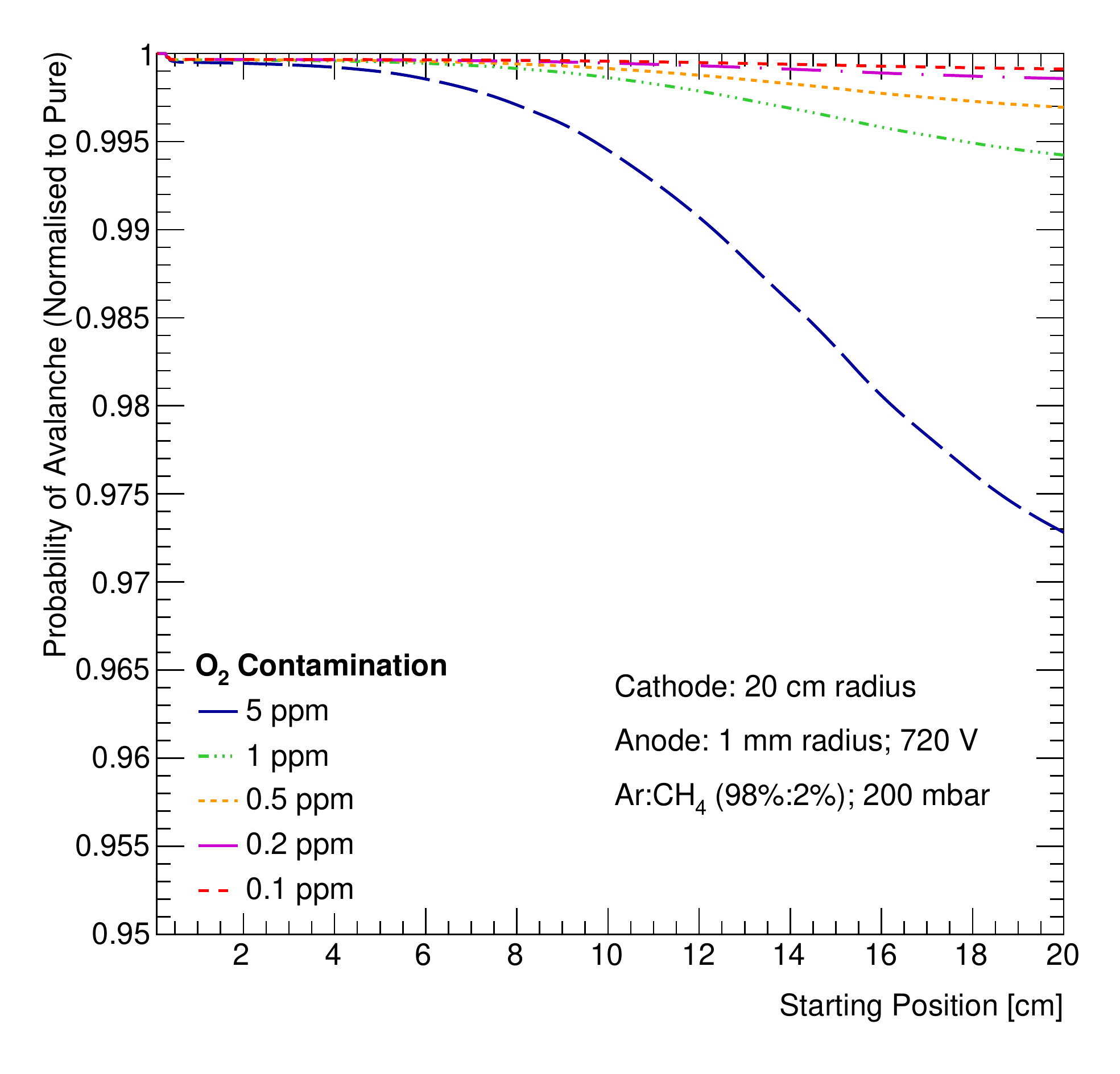}}
\subfigure[\label{fig:impurity2}]{\includegraphics[width=0.45\textwidth]{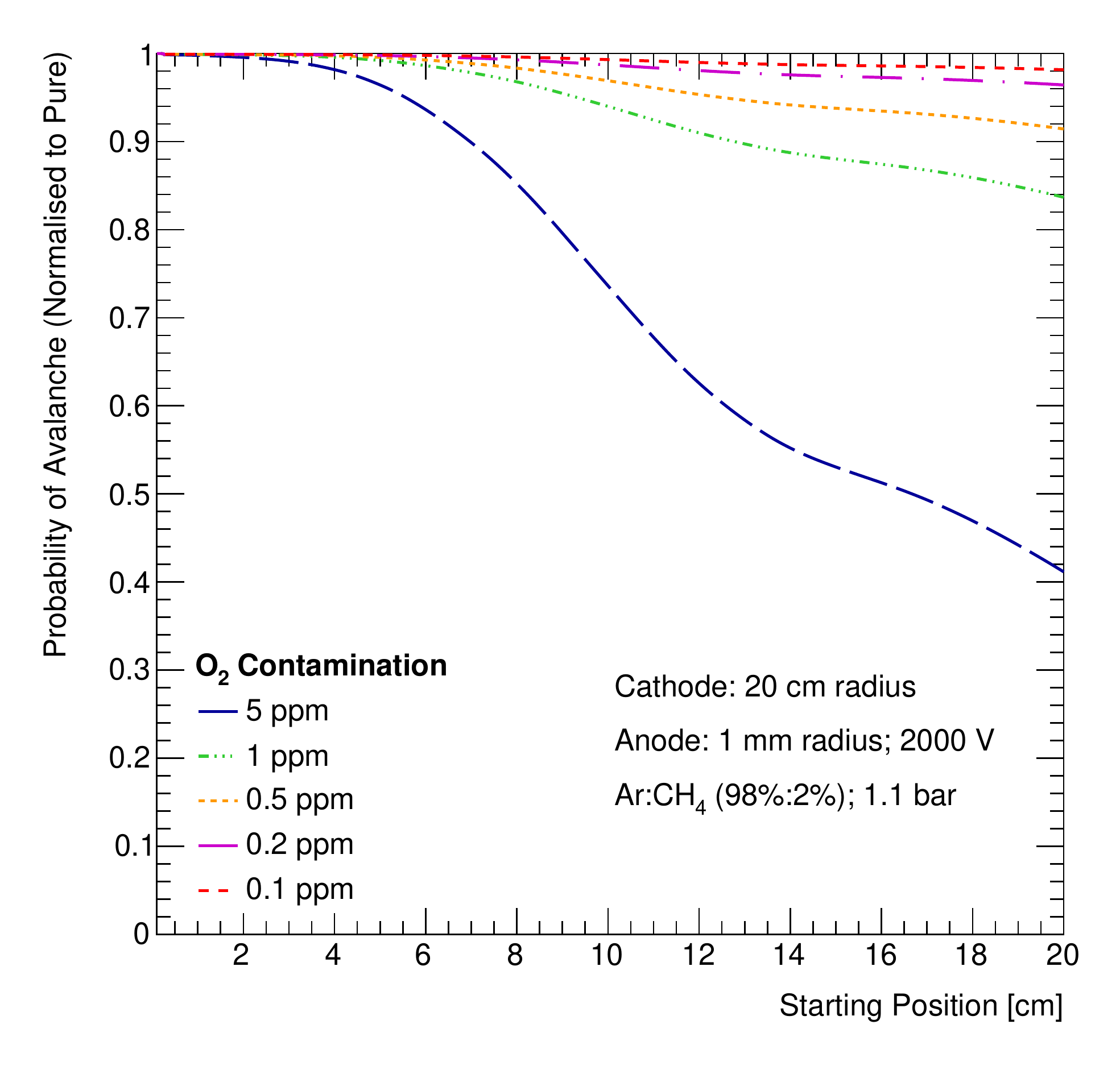}}
\caption{{\it The survival probability of an electron as a function of its
initial position for different levels of O$_2$ contaminations:
\subref{fig:impurity1} for Ar:CH$_4$ 98\%:2\% at 200~mbar and an anode voltage of 720~V, and
\subref{fig:impurity2} for the same gas mixture at a pressure of 1.1~bar and
anode voltage of 2000~V.\label{fig:impurity}}}
\end{figure}

This simulation is further used to investigate the properties
of the events, and complement the studies performed on the data. An initial application of the simulation to study the effect of electro-negative impurities is discussed.
The attachment
coefficient for 98\% Ar and 2\% CH$_4$ at 200~mbar with different levels of O$_2$ contamination
are presented in Fig.~\ref{fig:attachment}. The pure gas mixture  exhibits attachment in regions with large magnitude of the electric field, whereas
oxygen contamination results in attachment in regions with low magnitude of the electric
field. Due to the $1/r^2$ radial dependence of the electric field, the latter corresponds to the vast majority of the gas volume. The probability of an electron to initiate an
avalanche as a function of its 
initial position is shown in Fig.~\ref{fig:impurity1}, for different levels of impurities, at 200~mbar with anode
voltage at 720~V. The corresponding probability at 1.1~bar with anode
voltage at 2000~V is shown in Fig.~\ref{fig:impurity2}. These results highlight the importance of minimising impurities in the gas.


\section{Signal analysis}
\label{sec:signalTreatment}


The mean drift time of primary electrons 
depends on their production position, with typical drift times in the tens of $\mu$s.
%
During their drift, electrons are also subject to 
diffusion, with a standard deviation that increases with the cube of the drift distance,
which results in significant temporal dispersion.
%
Subsequently, electrons reach the high-field region,
approximately 1~mm from the anode, and the developed avalanche results in ions that slowly drift outwards, reaching the cathode surface in several seconds. The ion induced signal is reduced rapidly with radius,
and its majority is produced 
in the first few hundred $\mu$s of their drift.
However, the 150~$\mu$s time constant of the amplifier, in combination with the effects discussed earlier, leads to a significant ballistic deficit.
This is corrected for by removing the electronics response from the raw waveform through deconvolution.

\begin{figure}[t]
\centering
\includegraphics[width=0.75\textwidth]{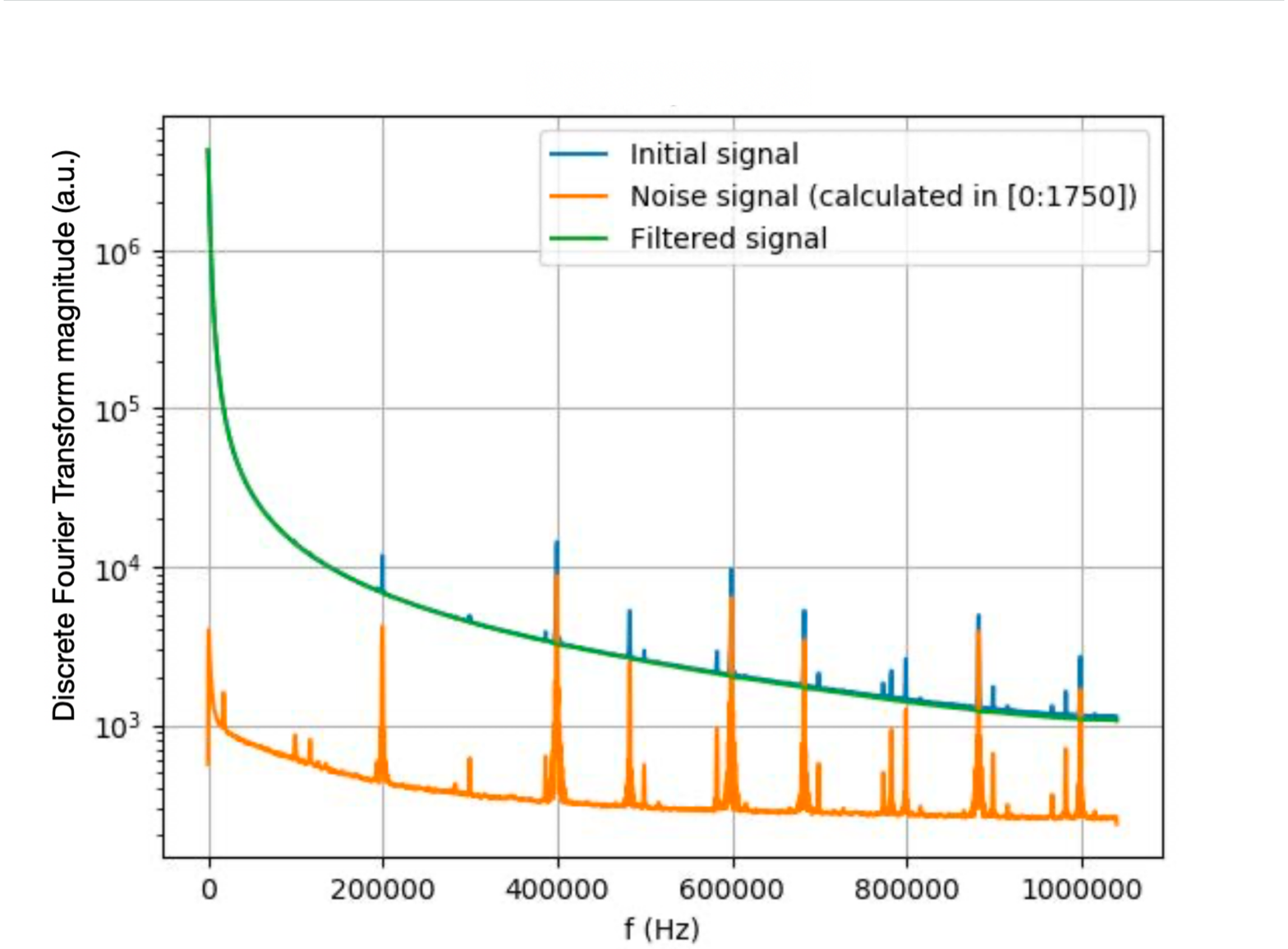} 
\caption{{\it Fourier transform of the average waveform before filtering (blue),  for the front parts of waveforms not containing a transient (orange), and after filtering (green).}\label{fig:7bis}}
\end{figure}

Initially, the raw
amplifier waveform $S(i)$ of 4166 samples, which is digitised at a sampling frequency $f_e$
of 2.08~MHz with a transient located at 50\% of the waveform length, is obtained and a baseline
correction to zero is applied.
In order to attenuate possible RFI, 
appropriate composite filtering is used~\cite{10.5555/281875,10.5555/524406} depending of the observed level of noise, as shown in Fig.\ref{fig:7bis}. The waveform is further 
 renormalised to the impulse response of the amplifier following:
\begin{equation}
S(i) = S(i)/(RC/t_e)
\end{equation}
where $t_e = 1/f_e$ and $RC/t_e = \sum_{n=0}^{n=\infty} e^{(-nt_e/RC)}$.

\begin{figure} [p]
\centering
  \subfigure[\label{fig:obs}]{\includegraphics[height=6.2 cm]{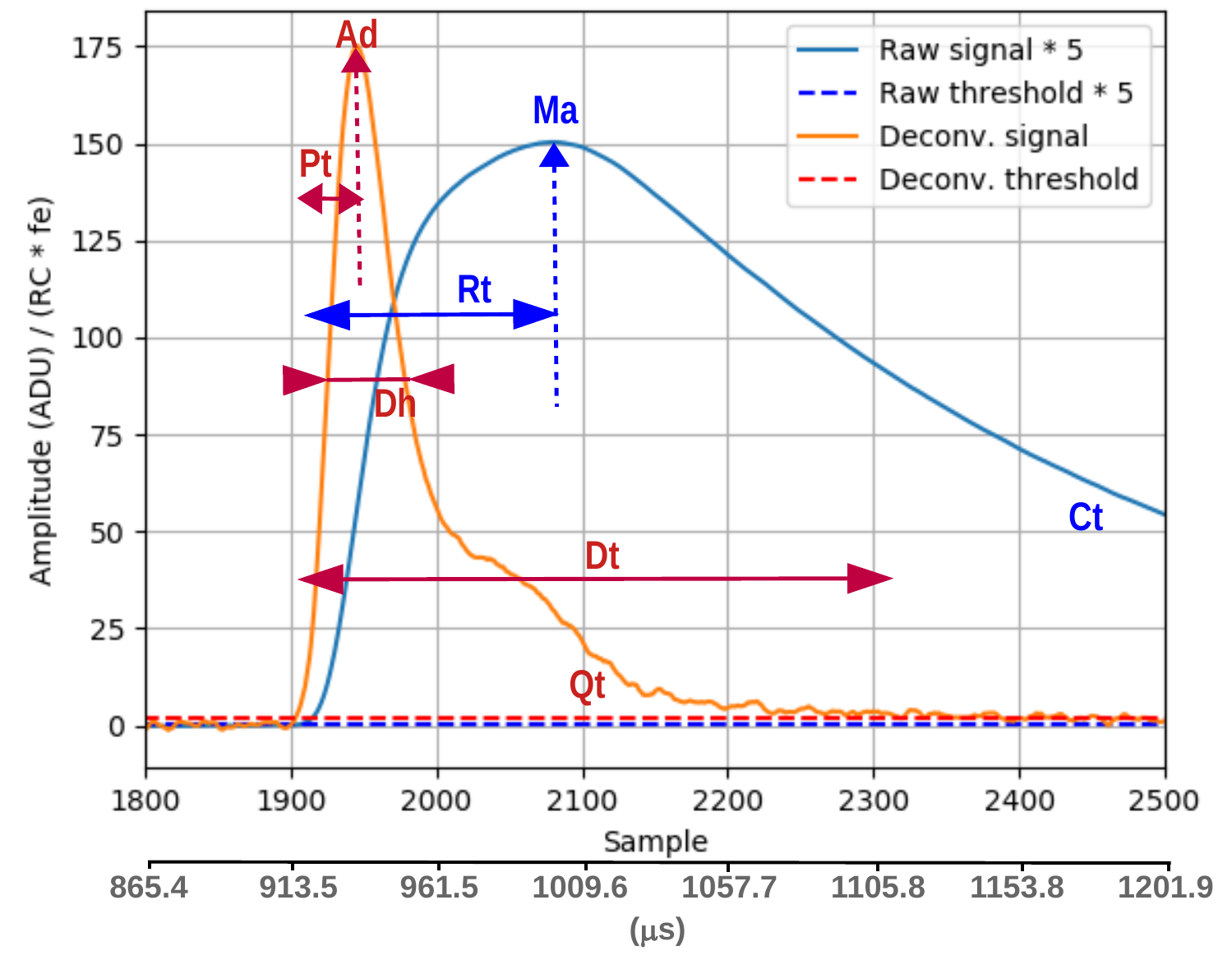}}
  \subfigure[\label{fig:sim_Pulses}]{\includegraphics[height=6.2cm]{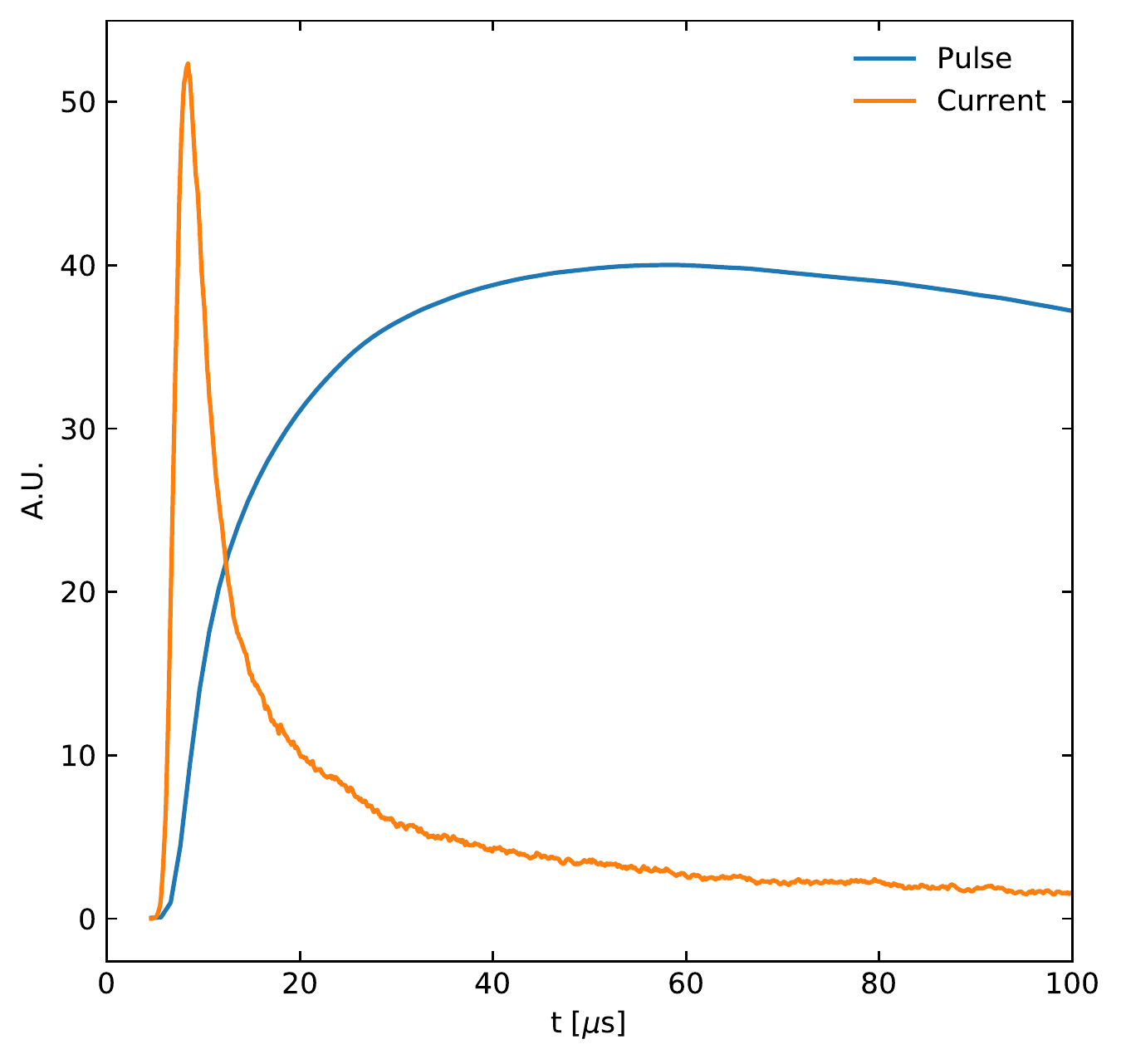}}
\caption{{\it  \subref{fig:obs} Real pulse with definition of observables used for the raw signal and for the deconvoluted one.
    \subref{fig:sim_Pulses} Example of simulated pulse, showing both the current and voltage signals.}}
        \label{fig:signals}
\end{figure}

The deconvolution is performed in the frequency domain, and as a cross-check also in the temporal
domain~\cite{JORDANOV1994592,nahman}.
For each 
waveform, a threshold is applied at 1\% of the waveform maximum amplitude and the first sample crossing the threshold defines the start for
the determination of the observables. 
An example is shown in Fig.~\ref{fig:obs}, while in Fig.~\ref{fig:sim_Pulses} a simulated pulse produced by a 5.3~MeV $\alpha$
particle is shown for comparison.
%
In the following, the variables related to
signal amplitude or integral are given in DAQ units (ADU) whereas the temporal
ones in $\mu$s. Three observables are extracted from the
raw signal:
\begin{itemize}
\item The pulse integral, 
$C_t$, between threshold crossing and last recorded sample.
\item The maximum amplitude, $M_a$.
\item The rise time, $R_t$, defined as time between  threshold crossing and maximum amplitude.
\end{itemize}

\begin{figure} [p]
\centering
\includegraphics[width=0.55\textwidth]{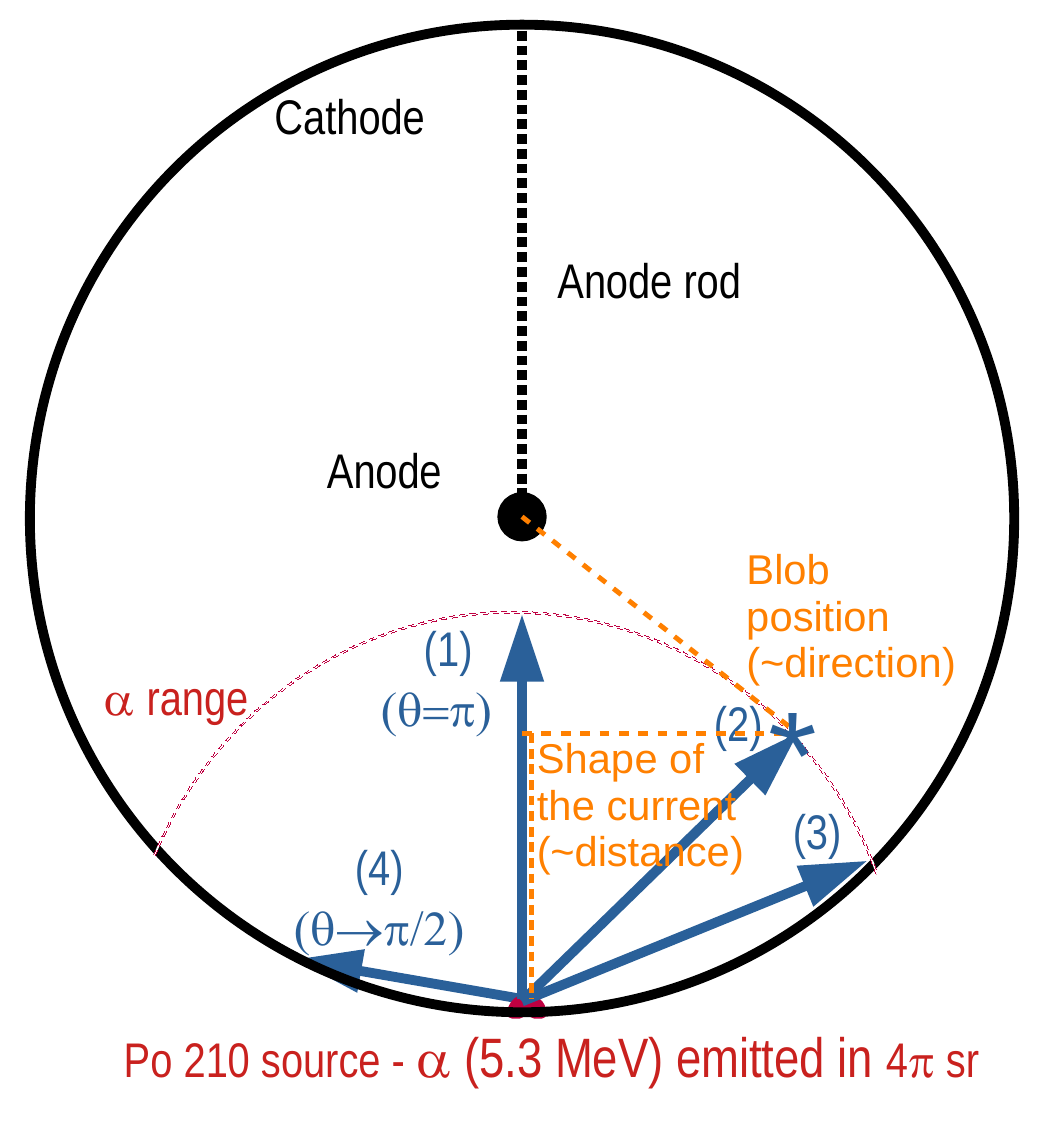}
\caption{{\it Interpretation of observables in relation to the different configurations of the tracks and geometry of the detector. The four typical configurations of tracks (labelled 1, 2, 3, and 4) correspond to $\cos(\theta)=-1, -0.8, -0.5, 0$ respectively.}}
\label{fig:obs1}
\end{figure}

For the deconvoluted signal,
four primary observables were constructed:
\begin{itemize}
\item The pulse integral, $Q_t$, between threshold crossing and last recorded sample.
This should equal $C_t$ and serves as check of the  deconvolution.
\item The pulse duration, $D_t$, defined as the time over threshold.
\item The pulse full width at half maximum, $D_h$.
\item The maximum amplitude, $A_d$.
\end{itemize}
From these observables, additional quantities are  derived to facilitate characterisation of the event:
\begin{itemize}
\item The peak time $P_t$,
giving the sample location of the signal maximum
$P_t(s)$, as a percentage of $D_t$, using : 
$P_t= P_t (s)/ D_t (s)$.
It
indicates the location of the maximum of signal in units of $D_t$, and gives the
direction of the track relative to the anode as shown in Fig.~\ref{fig:obs1}.
\item $D_i=Q_t/D_h$ aiming to quantify the temporal distribution of the primary electrons at their arrival at the anode, which could be sensitive to the average distance of the track to the cathode.

\end{itemize}

\begin{figure} [t]
\centering
  \subfigure[\label{fig:obsMa}]{\includegraphics[height=6.2
  cm]{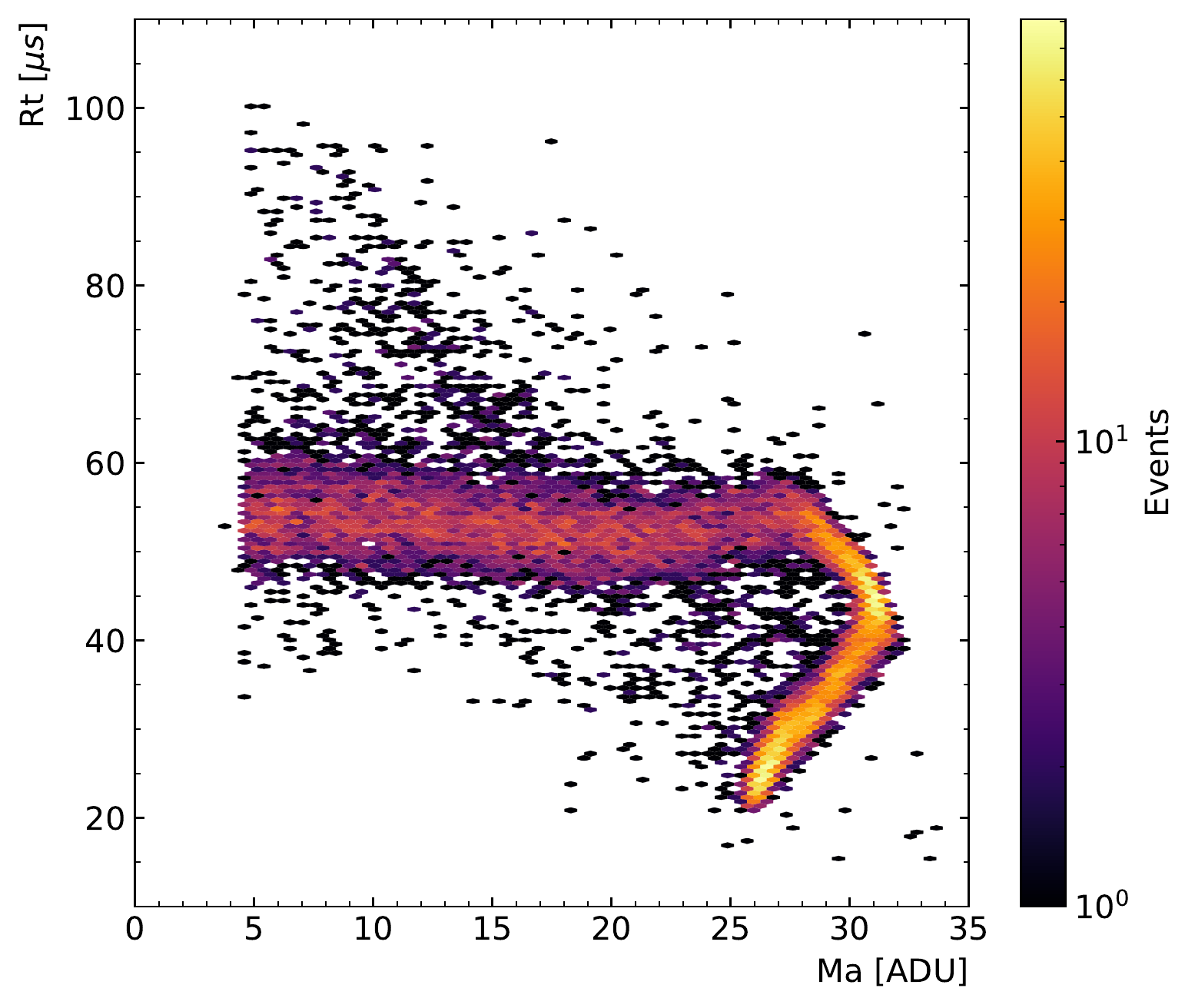}}
  \subfigure[\label{fig:simMa}]{\includegraphics[height=6.2cm]{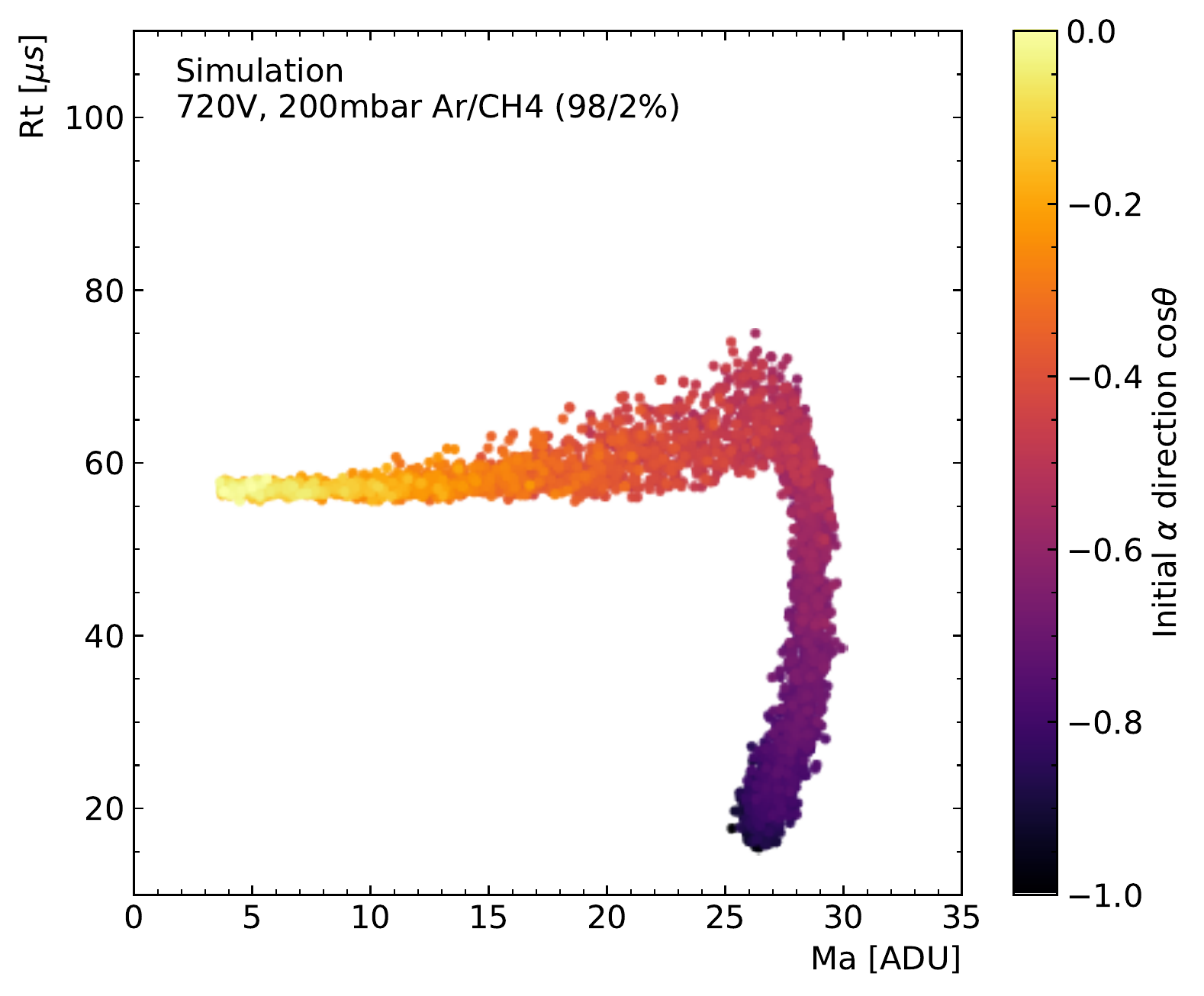}}
  \caption{{\it 2D distribution of rise time Rt versus maximal amplitude of raw
  signal Ma for \subref{fig:obsMa}~data and \subref{fig:simMa}~simulation. The
  ballistic deficit makes the Ma variable not suitable for reconstructing the
  $\alpha$ energy.}}
\label{fig:signalsMa}
\end{figure}
\begin{figure} [thp]
\centering
  \subfigure[\label{fig:matricsa}]{\includegraphics[height=5.5cm]{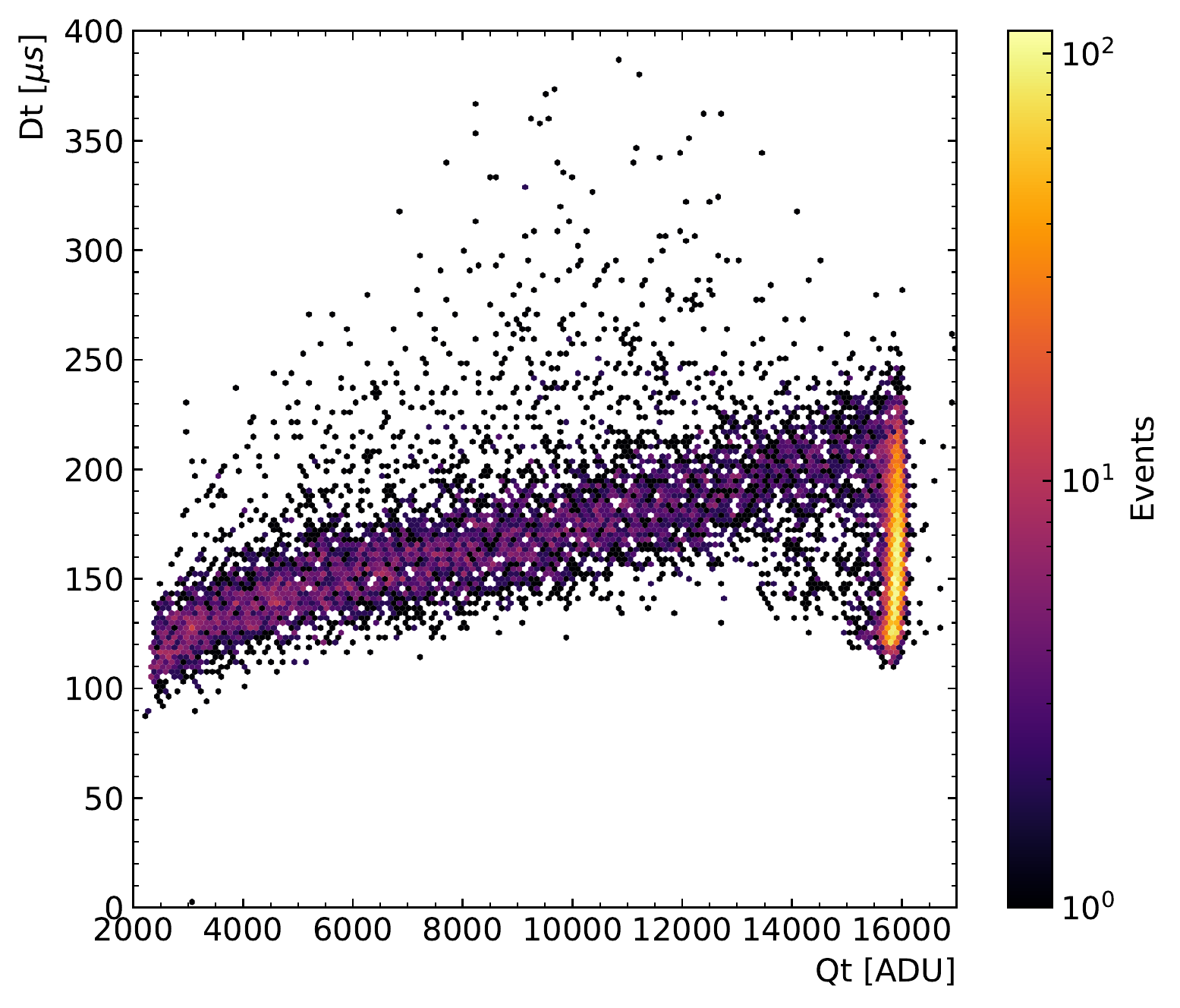}}
  \subfigure[\label{fig:matricsSimua}]{\includegraphics[height=5.5cm]{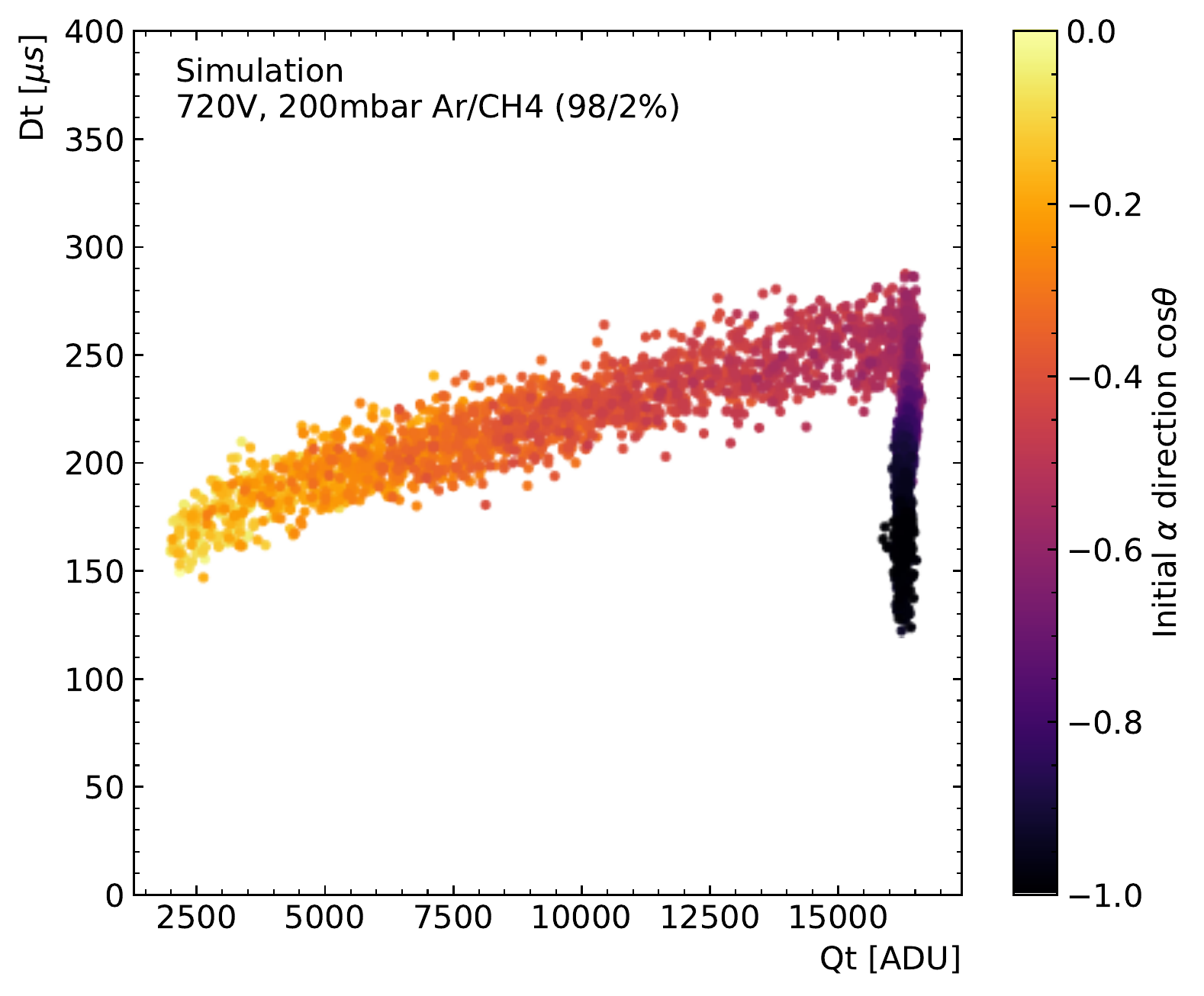}}
  \subfigure[\label{fig:matricsb}]{\includegraphics[height=5.5cm]{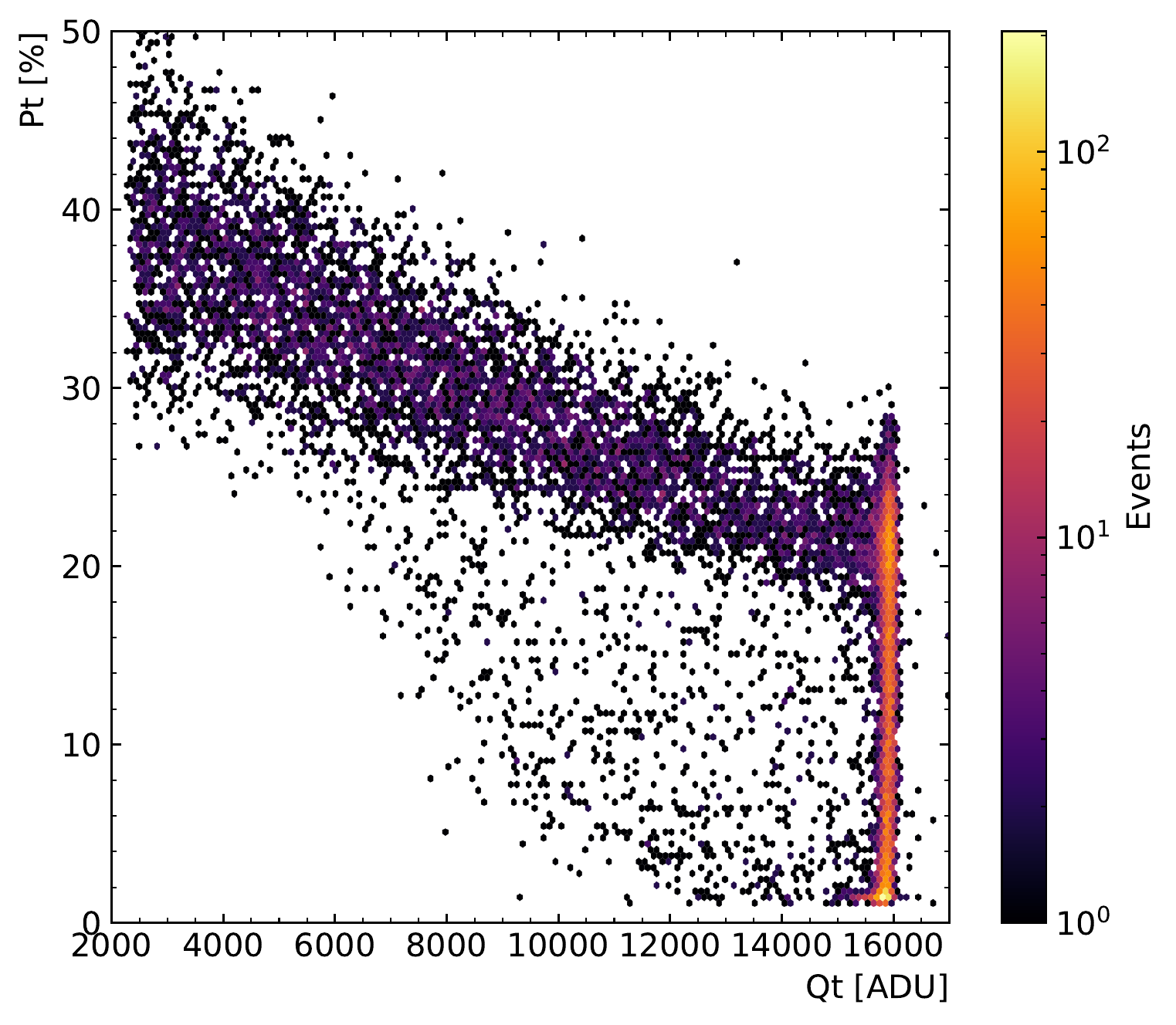}}
  \subfigure[\label{fig:matricsSimub}]{\includegraphics[height=5.5cm]{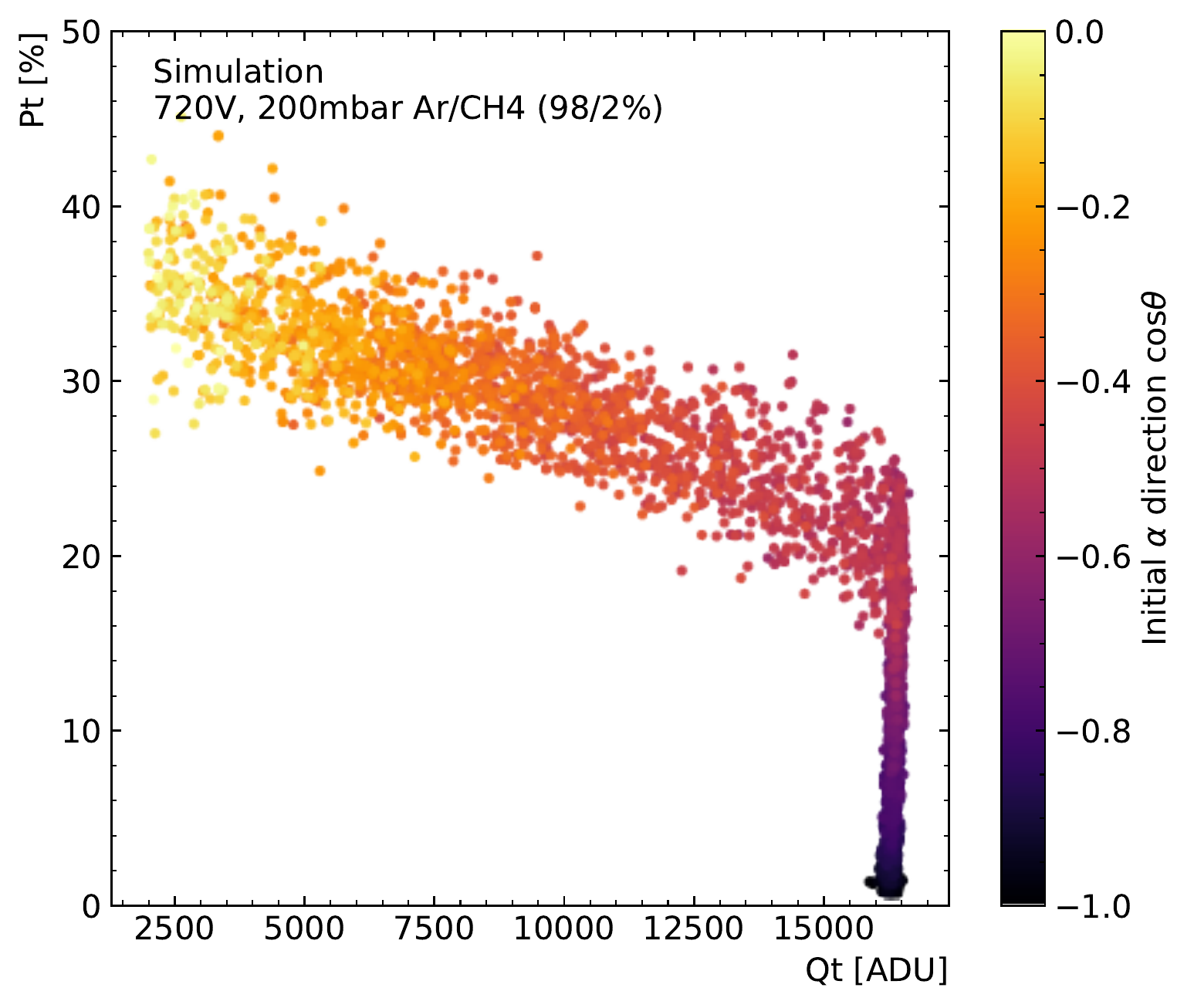}}
  \subfigure[\label{fig:matricsc}]{\includegraphics[height=5.5cm]{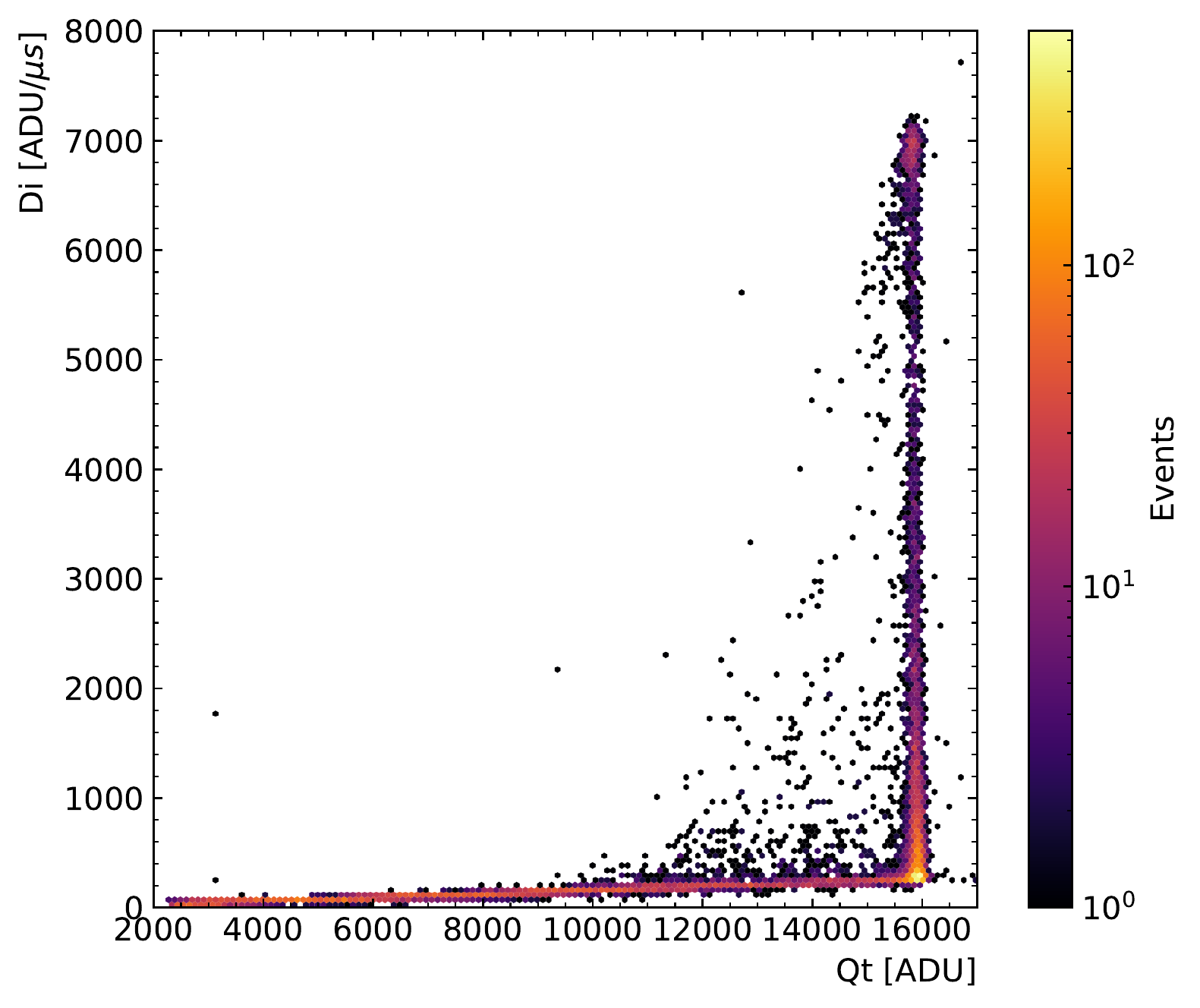}}
  \subfigure[\label{fig:matricsSimuc}]{\includegraphics[height=5.5cm]{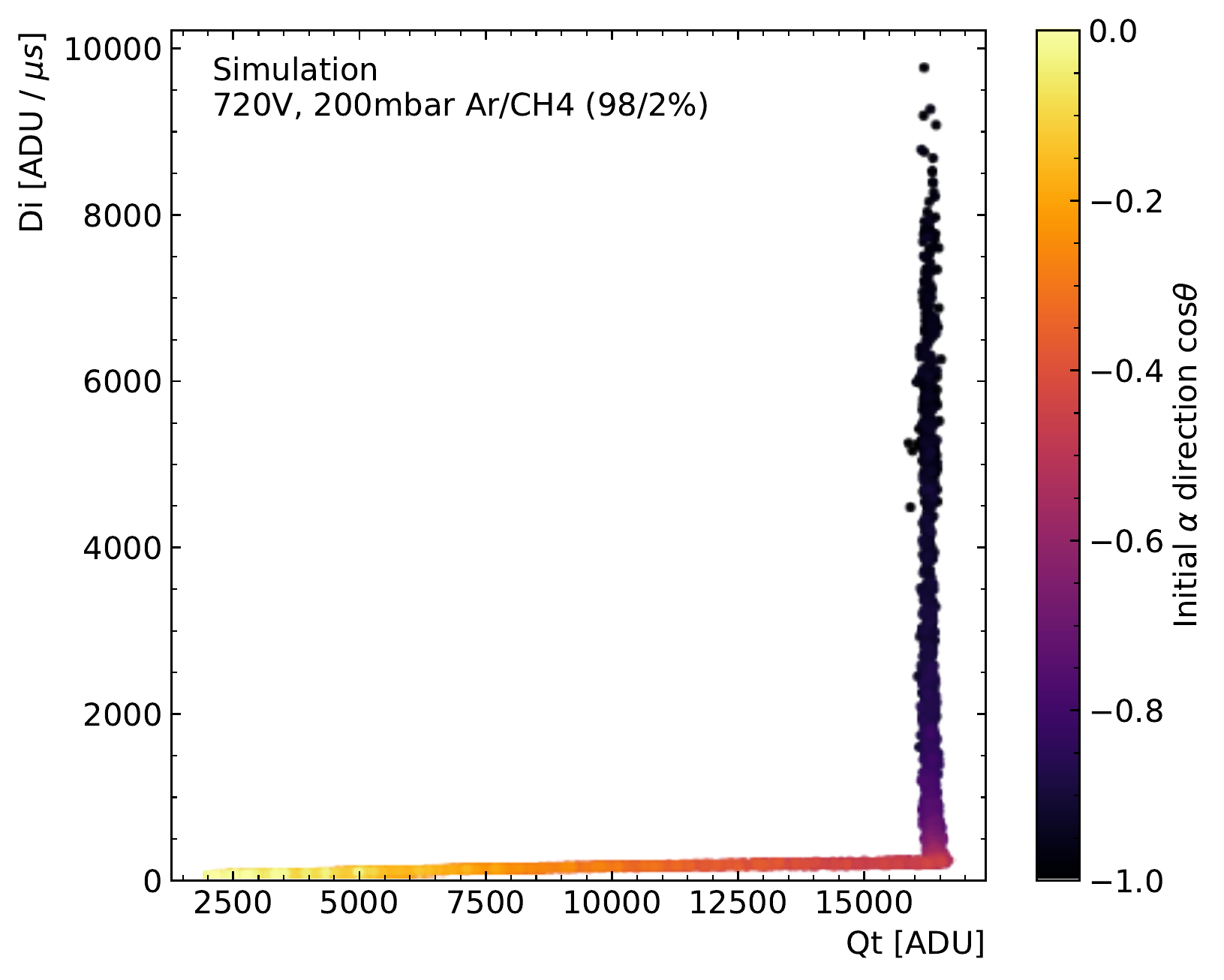}}
  \caption{{\it  Example of 2D representation of events for different
  observables for data at 200~mbar and 720~V (\subref{fig:matricsa}~Dt vs Qt, \subref{fig:matricsb}~Pt
  vs Qt, \subref{fig:matricsc}~Di vs Qt) and corresponding
  simulation  (\subref{fig:matricsSimua}~Dt vs Qt, \subref{fig:matricsSimub}~Pt vs
  Qt, \subref{fig:matricsSimuc}~Di vs Qt). For the
  data the colour indicates the number of events in the bin whereas for the
  simulation the colour indicates the original $\alpha$ direction: $\cos \theta =
  -1$ for tracks going towards the central anode and $\cos \theta = 0$ for
  tracks emitted orthogonally to the radial direction. The figures were made
  with about 14000 events without any selection cut, corresponding to about
  half an hour of data taking.}}
    \label{fig:matrics}
\end{figure}

In Fig.~\ref{fig:signalsMa} the $R_t$ versus $M_a$ are presented, which are the traditional variables used to analyse events~\cite{Arnaud:2018bpc}. In Fig.~\ref{fig:matrics}, the deconvoluted variables are shown, demonstrating that ballistic deficit is accounted for. These are directly compared with the simulation and a good agreement is observed.
As an example, 
in Fig.~\ref{fig:matricsa} the $Q_t$ variable  
does not depend on  $D_t$ for tracks fully contained in the gas volume, seen as a vertical accumulation at $Q_t$ of approximately 15500 ADU. Furthermore the track duration $D_t$ exhibits a dependence on the emission angle, as highlighted by the simulation in Fig.~\ref{fig:matricsSimua}. 

\begin{figure} [thp]
\centering
  \subfigure[\label{fig:data3d}]{\includegraphics[height=6cm]{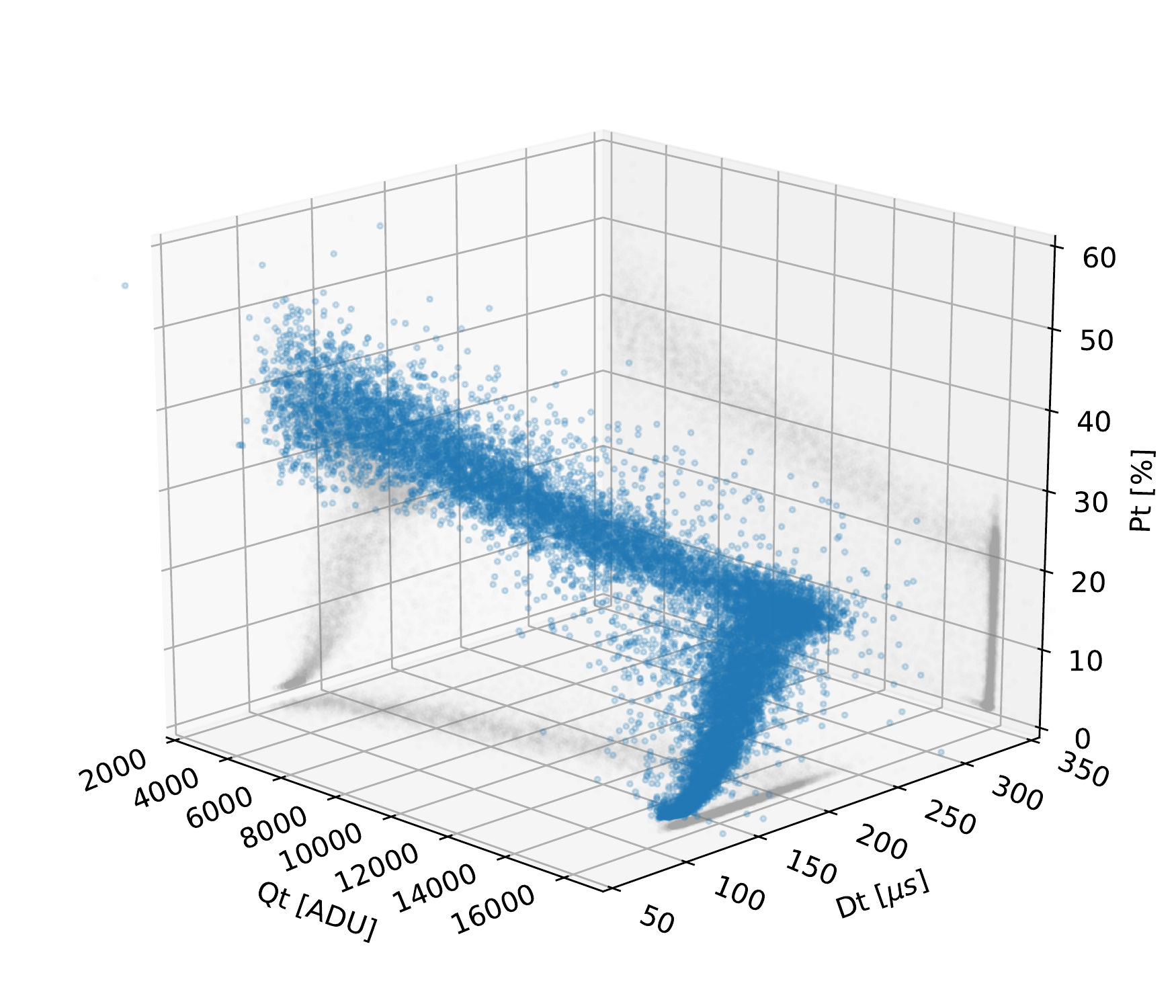}}
  \subfigure[\label{fig:simu3d}]{\includegraphics[height=6cm]{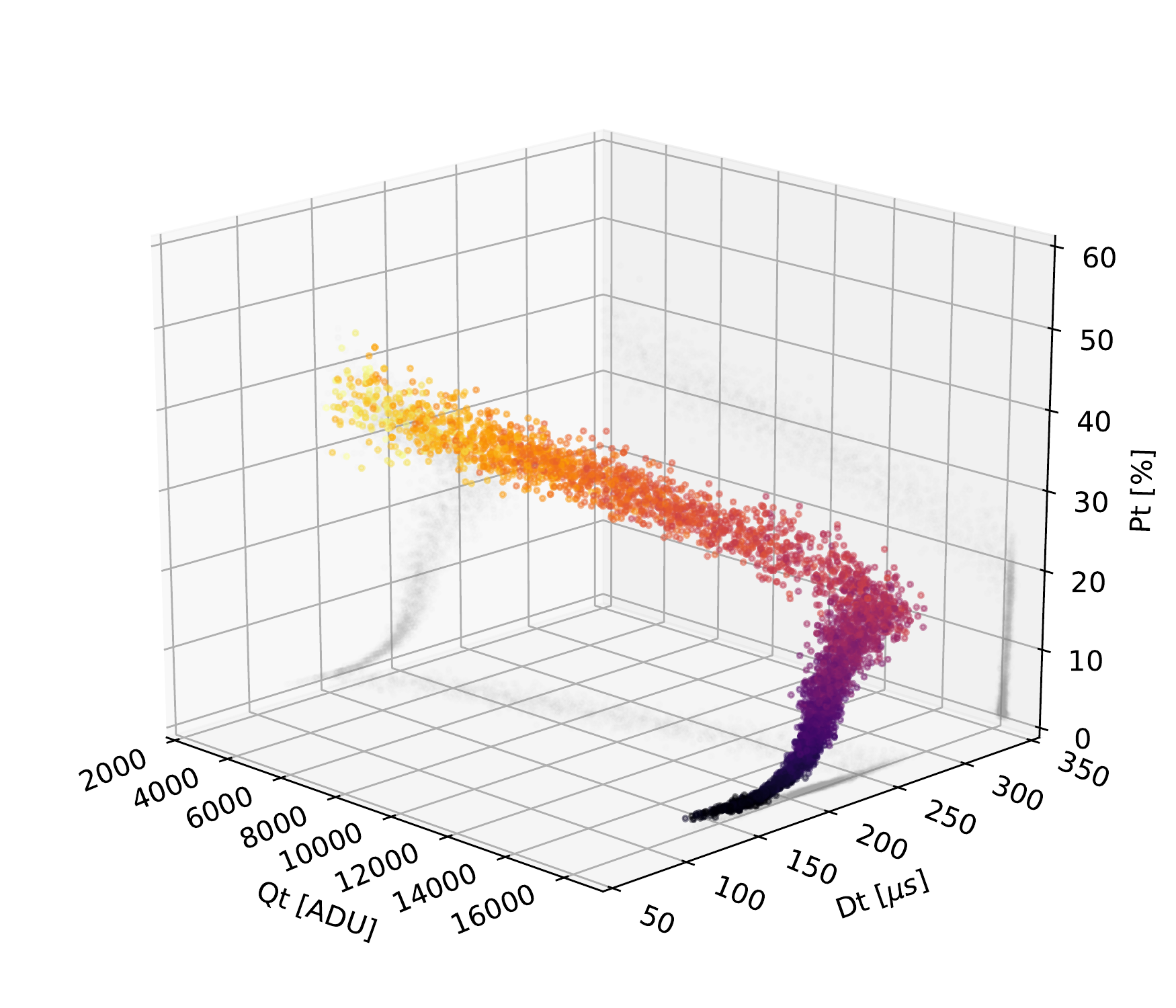}}
  \caption{{\it \subref{fig:data3d}~3D-plot of the observables (Qt, Dt, Pt)
  measured for the $^{210}$Po source at a pressure of 200~mbar.
  \subref{fig:simu3d}~The corresponding plot obtained from simulation,
  with the marker colours indicating the initial $\alpha$ direction, as in
  the previous figures.
  }}
\label{fig:sig3d}
\end{figure}
Some characteristics of long tracks can be inferred from the introduced observables. Specifically, the peak time  allows estimation of the track direction
 relative to the anode. For a track which is subject to a significant energy deposit at its end, i.e. Bragg peak, the deconvoluted signal must show a peak
corresponding to this deposit. Therefore, a track pointing towards the anode, corresponding to  configurations of
 tracks labelled 1 and 2 in Fig.~\ref{fig:obs1}, will present a maximum close to the beginning of the signal, leading to a $P_t$ in the range  0 to 30\%, as confirmed by the simulation in 
Fig.~\ref{fig:matricsSimub}.
In case of tracks hitting the cathode, for which the Bragg peak is therefore not present (e.g. configuration represented by label 4 in
Fig.~\ref{fig:obs1}.), the interpretation of $P_t$ in terms of track direction is more complex.
These cases point out the limits of this simple model.

%

The observable $D_i$, which is shown in
Figs.~\ref{fig:matricsc} and~\ref{fig:matricsSimuc}, can be interpreted by comparing to the simulation, where it could be 
related to the average distance of the track to the cathode and the shape of the current signal. 
For tracks stopping in the cathode, the current signals are bell-shaped with a $D_i$ slowly 
increasing with energy deposition.
This is no longer the case for tracks stopping inside the gas, below $\cos\theta = -0.6$, i.e. $Q_t>$15000 ADU and $D_i>$500 ADU/$\mu$s, for which the signal shape is dominated by the presence of a Bragg peak which breaks 
the bell-like symmetry of the signal. 

The main features of the presented understanding are checked by locating in the three-dimensional space $(Q_t, D_t, P_t)$ the events produced by the 
$^{210}$Po source.
As shown in Fig.~\ref{fig:sig3d}, these three observables could be exploited for the classification of
events inside the detector.
These observables offer a wide range of tools for event reconstruction and to probe various event configurations and topologies. Further dedicated studies will be required to demonstrate and quantify the potential for event discrimination and background suppression.
%


\section{Results}
\subsection{Detector stability}
\label{sec:decstability}
\begin{figure} [t]
\centering
\includegraphics[width=0.7\textwidth]{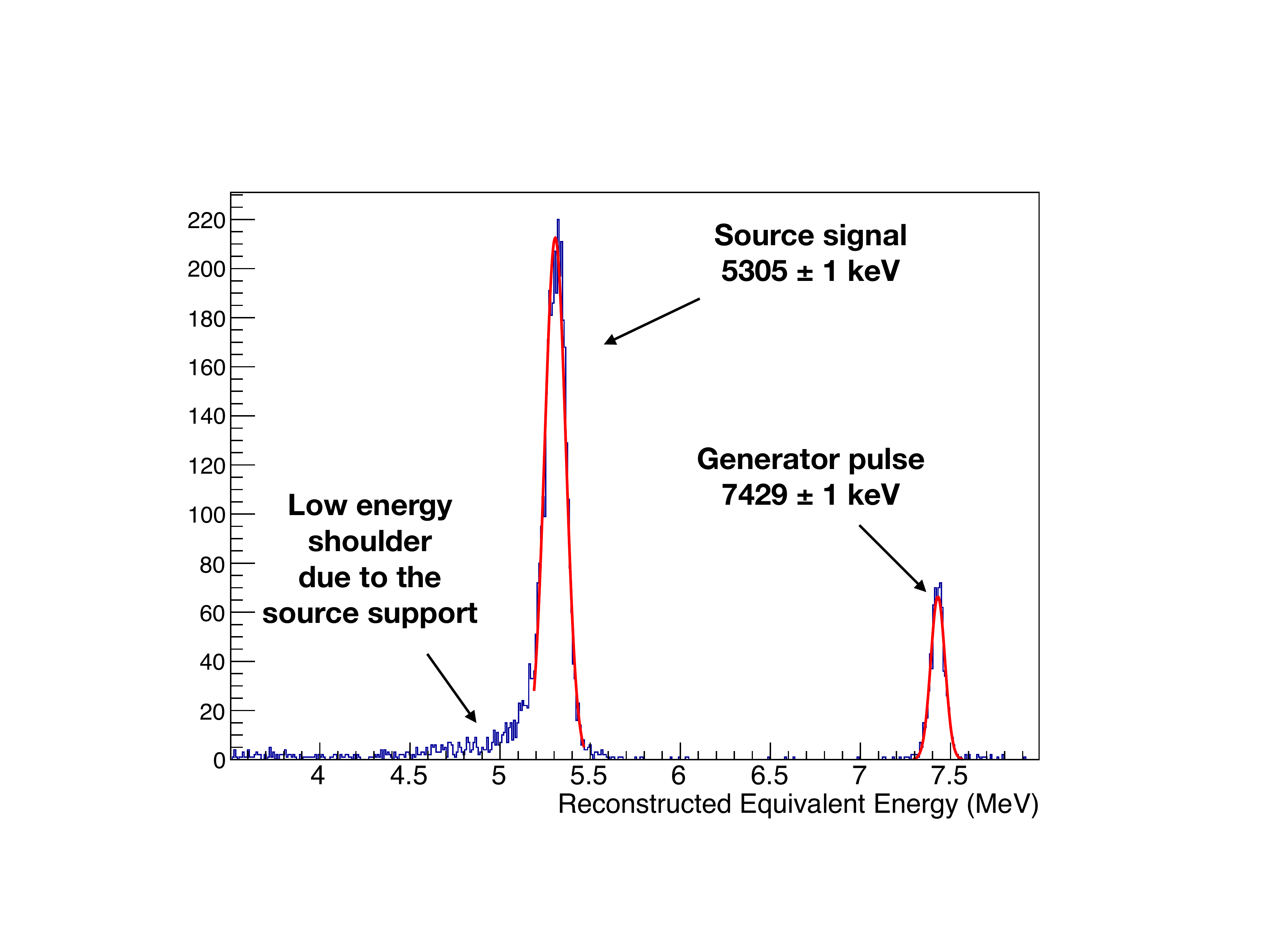}
\caption{{\it Example of the signal integral output for a run with $^{201}$Po source. The low energy shoulder due to the source support is clearly seen.}}
\label{fig:3}
\end{figure}
The detector stability in terms of gas gain was monitored using the $^{210}$Po $\alpha$ source. 
%
To disentangle detector effects from temperature or electronics response variations, in each run a square pulse was injected
in the test input of the pre-amplifier with frequency of
0.5~Hz. 

The first few hours after switching on the HV were discarded to allow the
detector to reach stable operation. Subsequently runs of 30 minutes each were recorded for a period of two weeks. The pulse integral was converted 
into equivalent reconstructed energy using the 5.305~MeV $\alpha$ peak of the $^{210}$Po source and assuming linear response, as
shown in Fig.~\ref{fig:3}.
These two peaks were fitted with Gaussian functions to
obtain the mean and its uncertainty.
These data 
were collected with a commercial CR-110-R2 CREMAT preamplifier~\cite{CREMAT}, with the addition of two protection diodes, 
and are not used to evaluate the energy resolution. 
%
No discharges were observed
and the preamplifier was eventually replaced by OWEN, described in 
Sec.~\ref{sec:elec}, with the advantage of having integrated protection diodes and reduced baseline fluctuations.
\begin{figure} [t]
\centering
\includegraphics[width=0.75\textwidth]{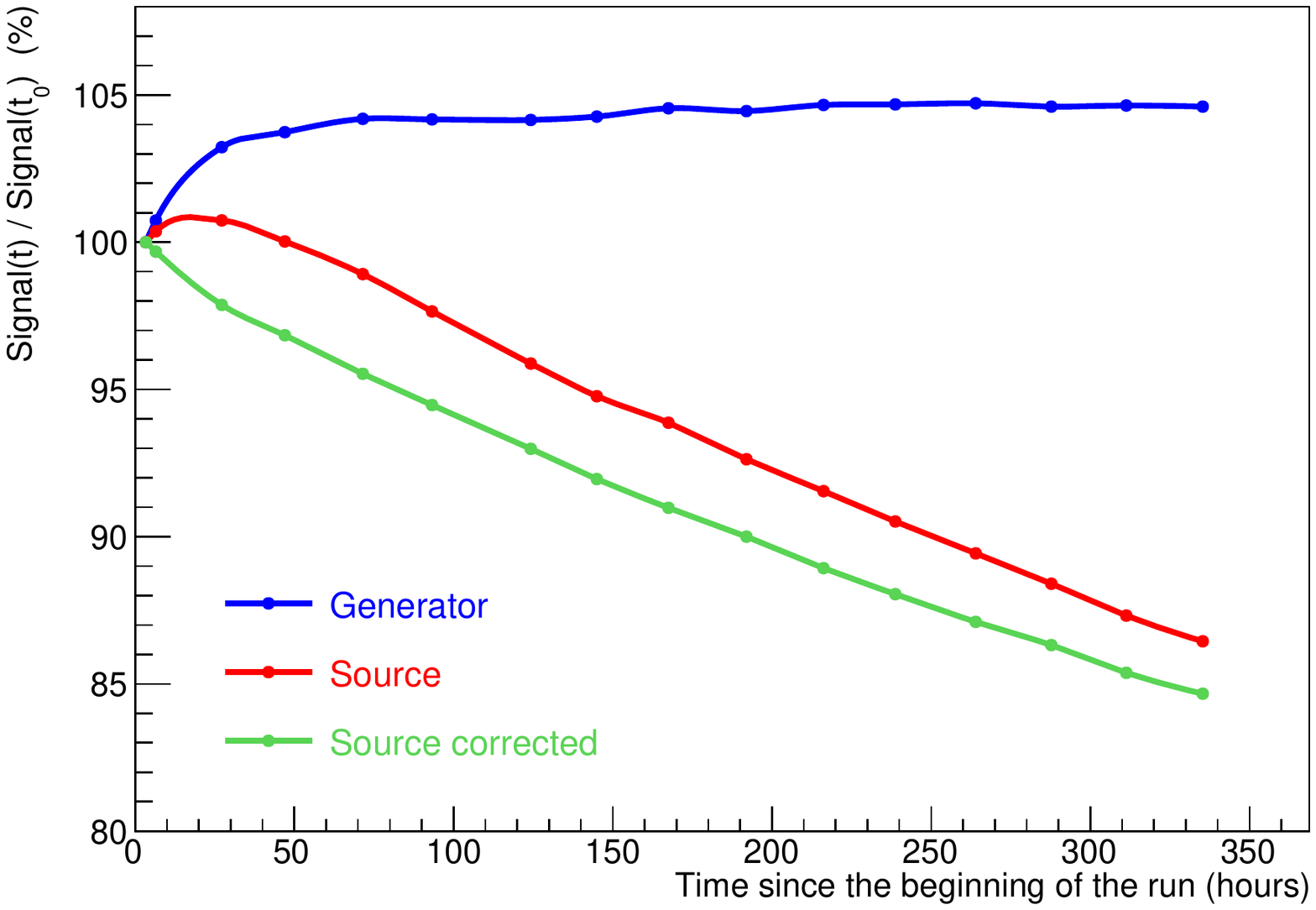}
\caption{{\it Time variation of the integral for generator pulses (blue line), $\alpha$ signal (red line) and $\alpha$ signals corrected according to the generator pulse value (green line).}}
\label{fig:4}
\end{figure}
The signal integral as a function of time is shown in Fig.~\ref{fig:4}. The generator pulses are rather stable as expected (blue line), nonetheless for the first two days an increase of the signal is still observed. 
Similar behaviour is observed for $\alpha$ signals (red line). The pulse generator data allow to disentangle   variations of the gas gain, by correcting for electronic gain variations. 

Once the signal is corrected for electronic gain variation, a gas gain loss as a function of time is observed. This is expected and it amounts to approximately $0.05\%/h$. This can be corrected offline and allows for good quality data taking
over several weeks. Furthermore, this effect can be suppressed further by reducing leaks and
materials outgassing, and by recirculating the gas through an oxygen removal cartridge.


\subsection{Energy resolution}
\label{sec:EnergyResolution}

\begin{figure} [t]
\centering
\includegraphics[width=0.75\textwidth]{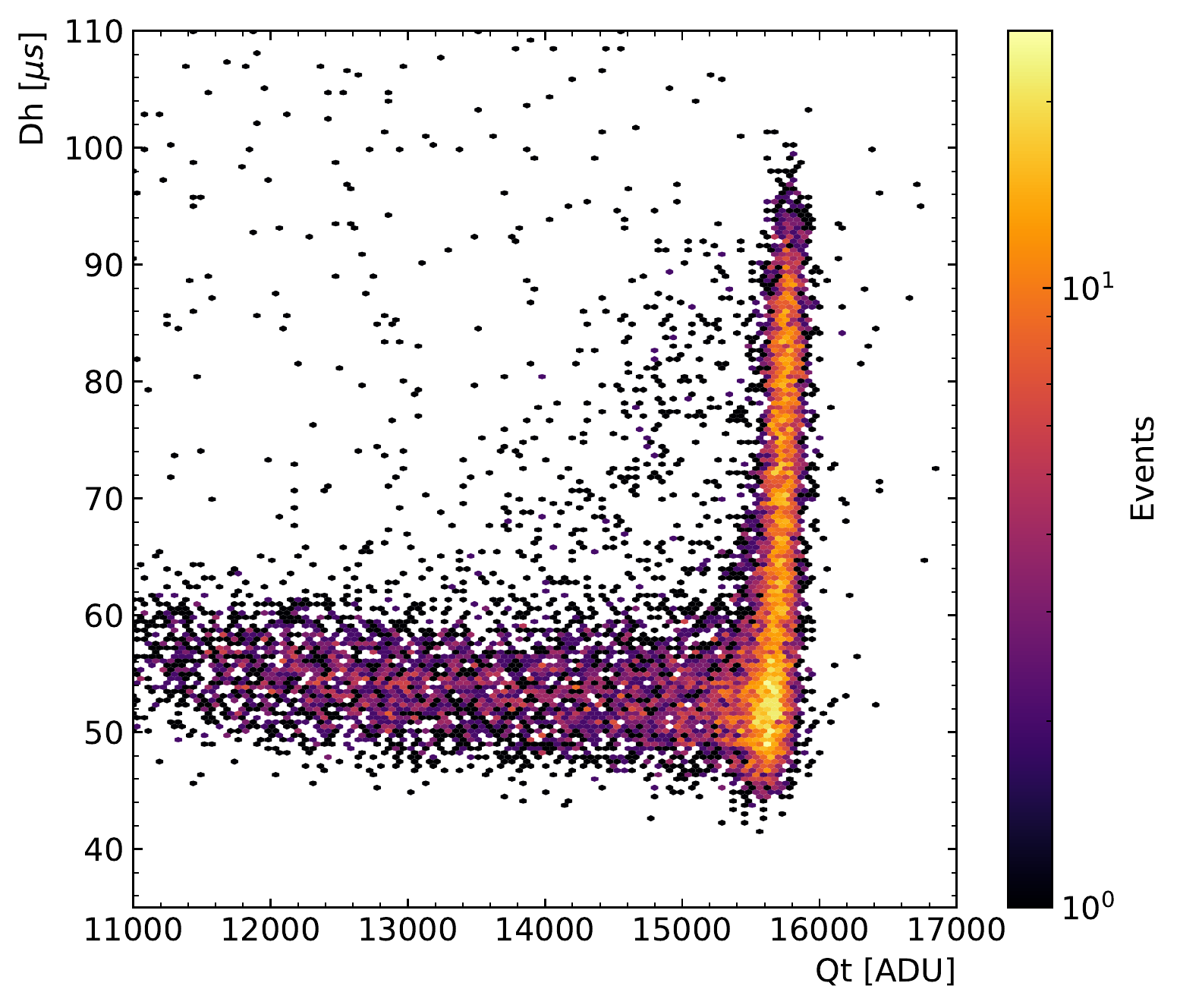}
\caption{{\it Two dimensional plot of signal width at half maximum Dh versus the measured charge Qt of the 5.3~MeV $\alpha$ signal at 1.1~bar and 2000~V.}}
\label{fig:7}
\end{figure}

The OWEN preamplifier was used to estimate the energy resolution.
A baseline noise, quantified as the RMS 
variation of the baseline in a single event, 
of about 8.2 ADU, and a baseline stability, defined as the RMS of the mean baseline of all events, of 13.1 ADU, were obtained. 
The width of the reconstructed pulse generator
signal corresponds to an intrinsic energy resolution of 0.5\% FWHM at the
$\alpha$ peak. This is driven by the electronic chain -- preamplifier,
DAQ, HV -- and may be the ultimate limiting factor. The OWEN project aspires, among other things, to develop a low noise readout to further reduce this contribution. 

\begin{figure} [t]
\centering
  \subfigure[\label{fig:6a}]{\includegraphics[width =
  0.49\textwidth]{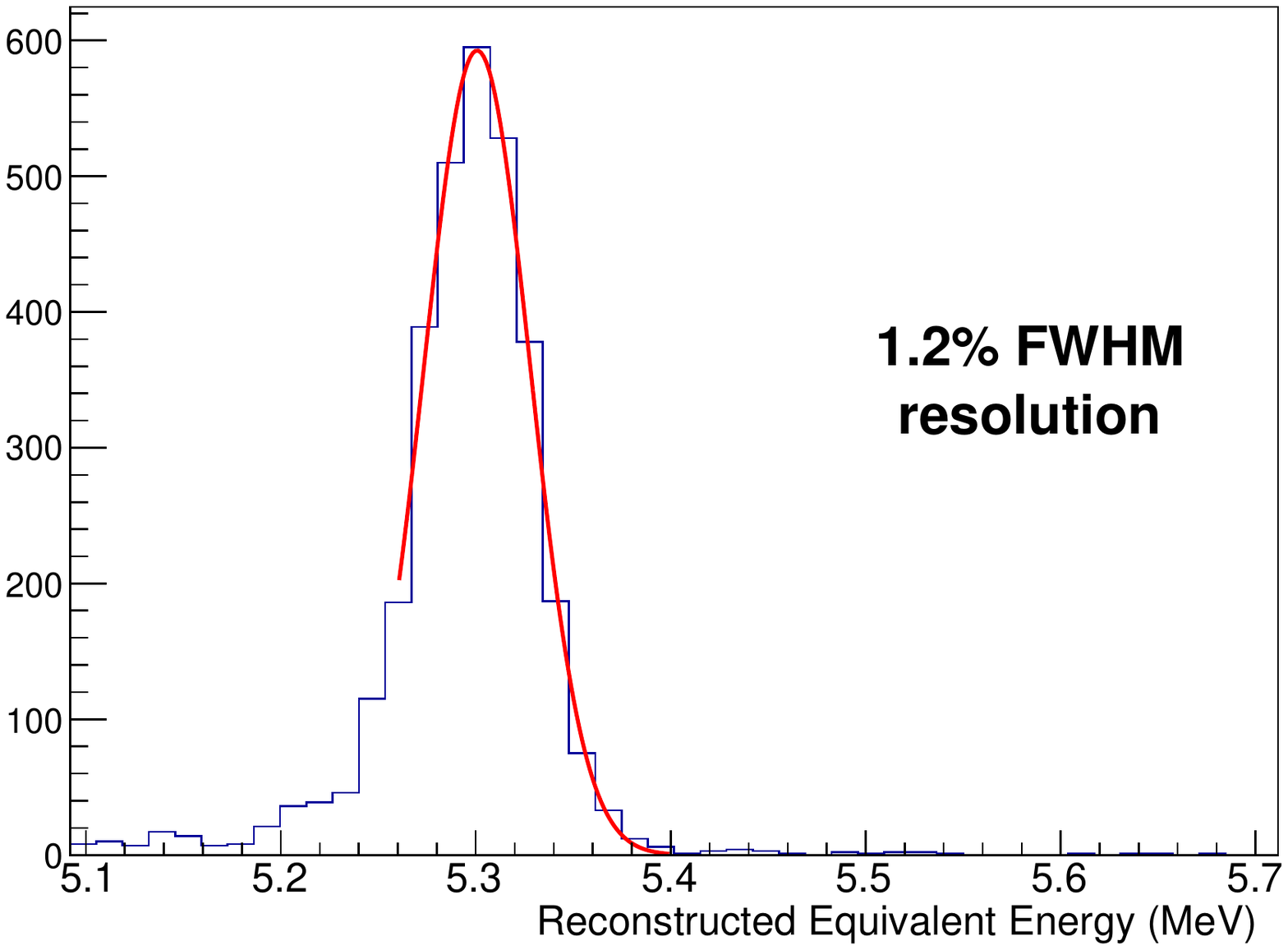}}
  \subfigure[\label{fig:6b}]{\includegraphics[width =
  0.49\textwidth]{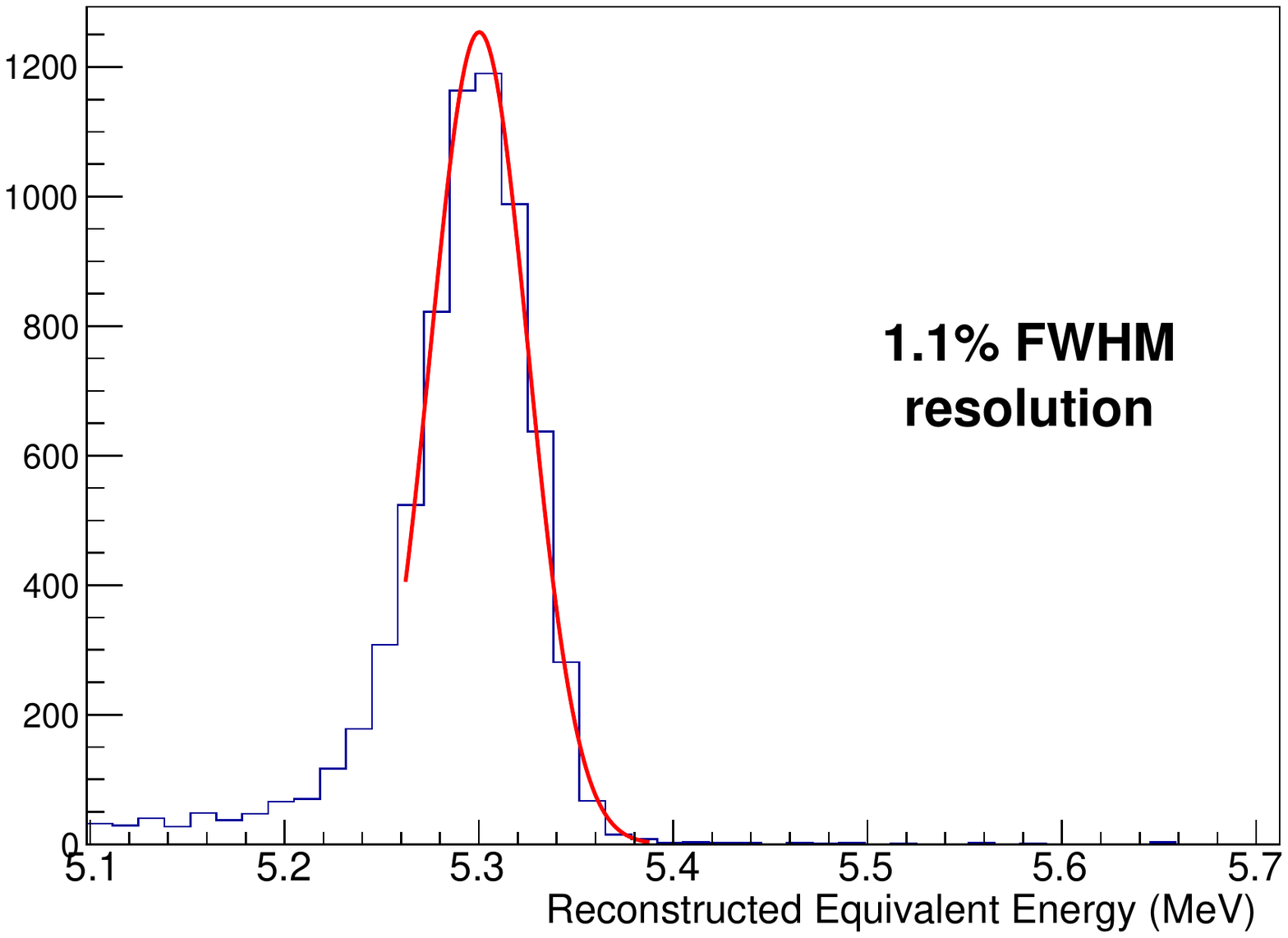}} \caption{{\it Integral of the 5.3~MeV
  $\alpha$ signal at \subref{fig:6a}~1.1~bar and 2000~V and at
  \subref{fig:6b}~200~mbar and 720~V. The Gaussian fit in red shows an energy
  resolution of 1.2\% FWHM and 1.1\% FWHM respectively.}}
\label{fig:6}
\end{figure}

The signal treatment presented in
Sec.~\ref{sec:signalTreatment} was applied to data
taken at 1.1~bar with a voltage of 2000~V on the central anode.
Events are selected with a pulse width at half maximum, $D_h > 65~\mu$s to remove 
the low energy tail seen Fig~\ref{fig:7}.
As shown in Fig.~\ref{fig:6a}, the reconstructed $Q_t$ exhibits a resolution of 1.2\%~FWHM. 
The low energy
tail, arising from events depositing energy in the source support, has been excluded from the fit. This resolution includes a smearing due to the source
itself, estimated to about 0.4\%, and for possible inhomogeneities on the
central anode (i.e. different gain depending on the starting point of the
ionisation electrons). 


The energy resolution dependence 
on track length is important as electrons emitted in $\beta\beta$ decays are expected to leave long tracks. 
This is studied by collecting data at a pressure of 200 mbar.
An anode HV of 720~V was used, providing the same gain as obtained in the 1.1 bar dataset. At this pressure $\alpha$ particles leave
tracks of approximately $20$~cm, a five-fold increase with respect to 1.1 bar.  An improved energy resolution of 1.1\% FWHM is obtained, as shown in Fig.~\ref{fig:6b}, demonstrating a robust energy resolution as a function of the track length.


By statistically subtracting the contributions of the read-out electronics and the source support, an energy resolution resolution of 0.97\% is obtained. 
%
Scaling the measured energy resolution of 1.1\% at 5.3~MeV to the $Q_{\beta\beta}$ of 2.458~MeV,
 assuming a $\sqrt{E}$ dependence, yields a resolution of 1.6\%.
%
It is noted that 
for Ar:CH$_4$ 98\%:2\%, the expected intrinsic energy resolution due to stochastic fluctuations of the number of generated electrons-ion pairs is estimated to be 0.24\%.
These results are expected to further improve for xenon, where the W-value is lower than argon~\cite{PhysRev.103.1253,Aprile:2009dv,Gomez-Cadenas:2019ges}.
The excellent energy resolution achieved and the demonstration of the independence to track length are important milestones for the  R2D2 project and a future $\beta\beta0\nu$ search with a spherical TPC.



\subsection{Discussion on the obtained results}
\label{sec:results_discussion}

\begin{figure}[t]
\centering
\includegraphics[width=0.75\textwidth]{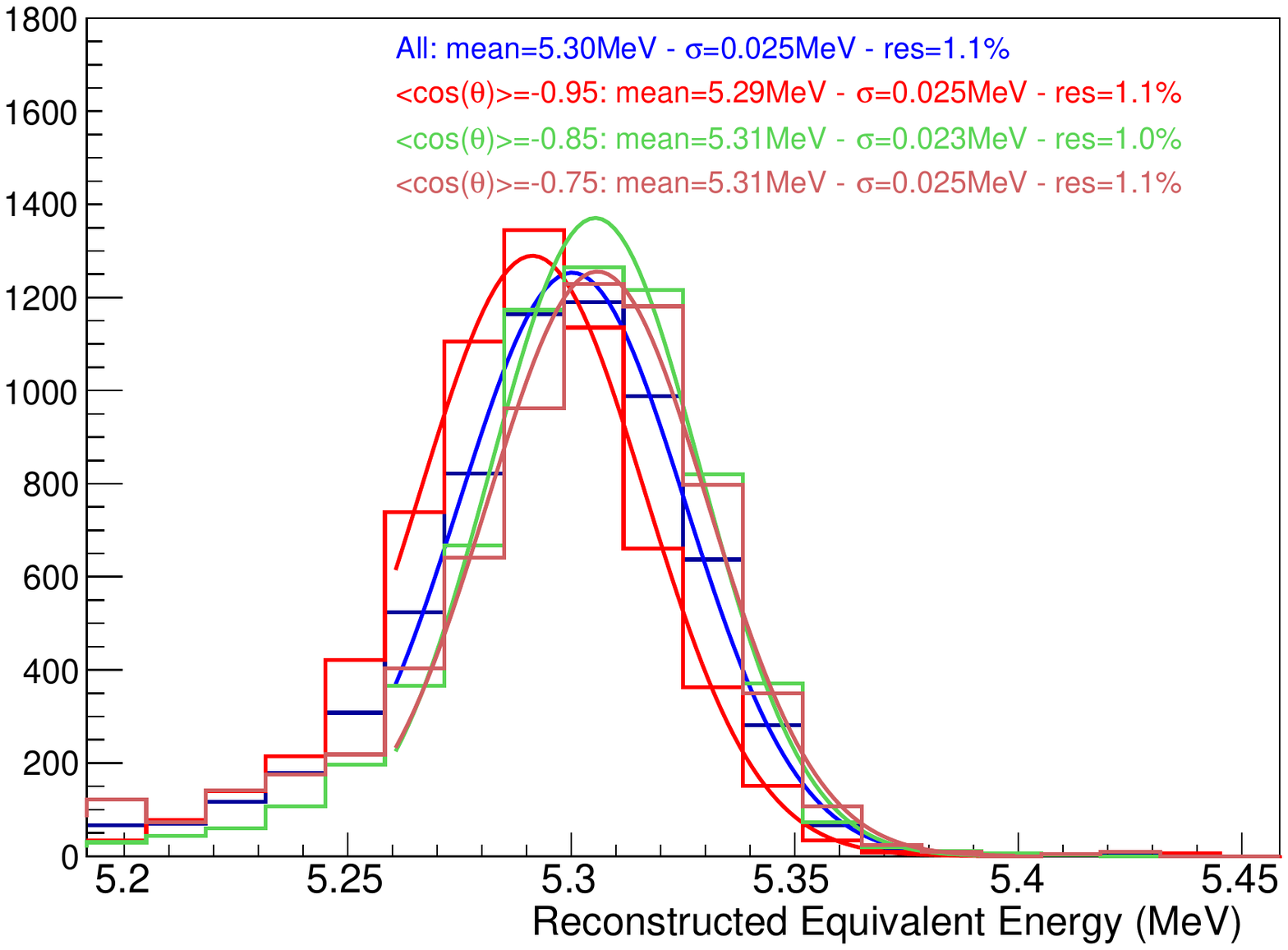}
\caption{{\it Integral of the 5.3~MeV $\alpha$ signal at 200~mbar and 720~V for different alpha angle emission: all events (blue), mean cos($\theta$) equal to -0.95 (red), -0.85 (green) and -0.75 (brown). The samples corresponding to a specific angle were normalized to the total statistics. The Gaussian fits are also shown.}}
\label{fig:18}
\end{figure}

%
%
To investigate the energy resolution in different regions of the 
gas volume, the 200~mbar dataset was categorised based on $D_t$, effectively selecting on $\alpha$ emission angles, as shown in Fig.~\ref{fig:matricsSimua}.
Three regions
were selected: 120 -- 150 $\mu$s, 150 -- 180 $\mu$s  and
180 -- 210 $\mu$s, corresponding to average $\cos\theta$ values of approximately -0.95, -0.85, and -0.75, respectively.
%
For angles smaller than $\cos\theta \approx -0.6$ the $\alpha$ particles 
only partially deposit their energy in the detector before hitting the wall. The reconstructed energy distributions for the three categories and for the inclusive selection
are shown in Fig.~\ref{fig:18}. Compatible energy resolutions are obtained in these three regions, demonstrating independence with respect to the track direction.
The relative variation in the mean 
reconstructed energy for the different
 regions is 0.37\%, well below the estimated energy resolution.
%

\begin{figure} [t]
\centering
\includegraphics[width=0.75\textwidth]{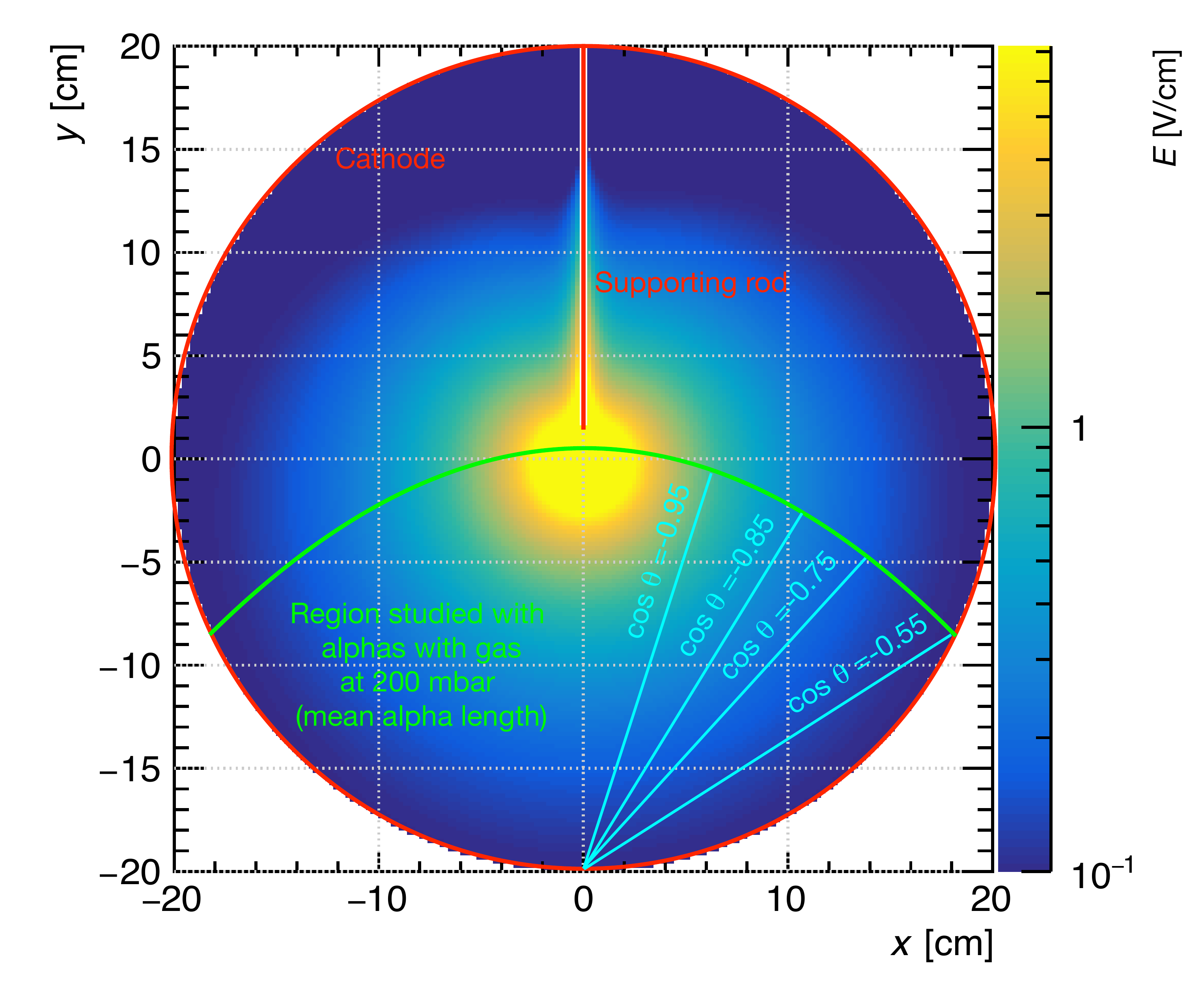}
\caption{{\it Electric field in V/cm as a function of x and y for z=0 for 200 mbar gas with a central anode at 720~V. Given the 1/r$^2$ dependence, the field very close to the central anode reaches values up to tens of kV/cm, however in the figure the colour scale is saturated at 7 V/cm in order to clearly show the field variation in the region of interest for the considered alpha tracks. The red line represent the sphere whereas the green line limits the region investigated using alpha tracks of 5.3~MeV considering the mean alpha length. The light blue lines indicates some specific alpha angles used in the analysis detailed in the text.}}
\label{fig:17}
\end{figure}
The $\alpha$ tracks used in this analysis are fully contained in the
bottom hemisphere of the detector, far from the anode support rod. In this region the electric field is to a good approximation that of an ideal spherical detector, as shown in Fig.~\ref{fig:17}.
Distortion of the electric field by the supporting rod is a known effect, and a number of approaches are proposed to mitigate this, e.g. applying a correction voltage~\cite{Katsioulas:2018pyh,Giomataris:2020rna}. This is further discussed in Sec.~\ref{sec:improvements}. 
%
%
%

\begin{figure}
    \centering
    \includegraphics[width=0.55\textwidth]{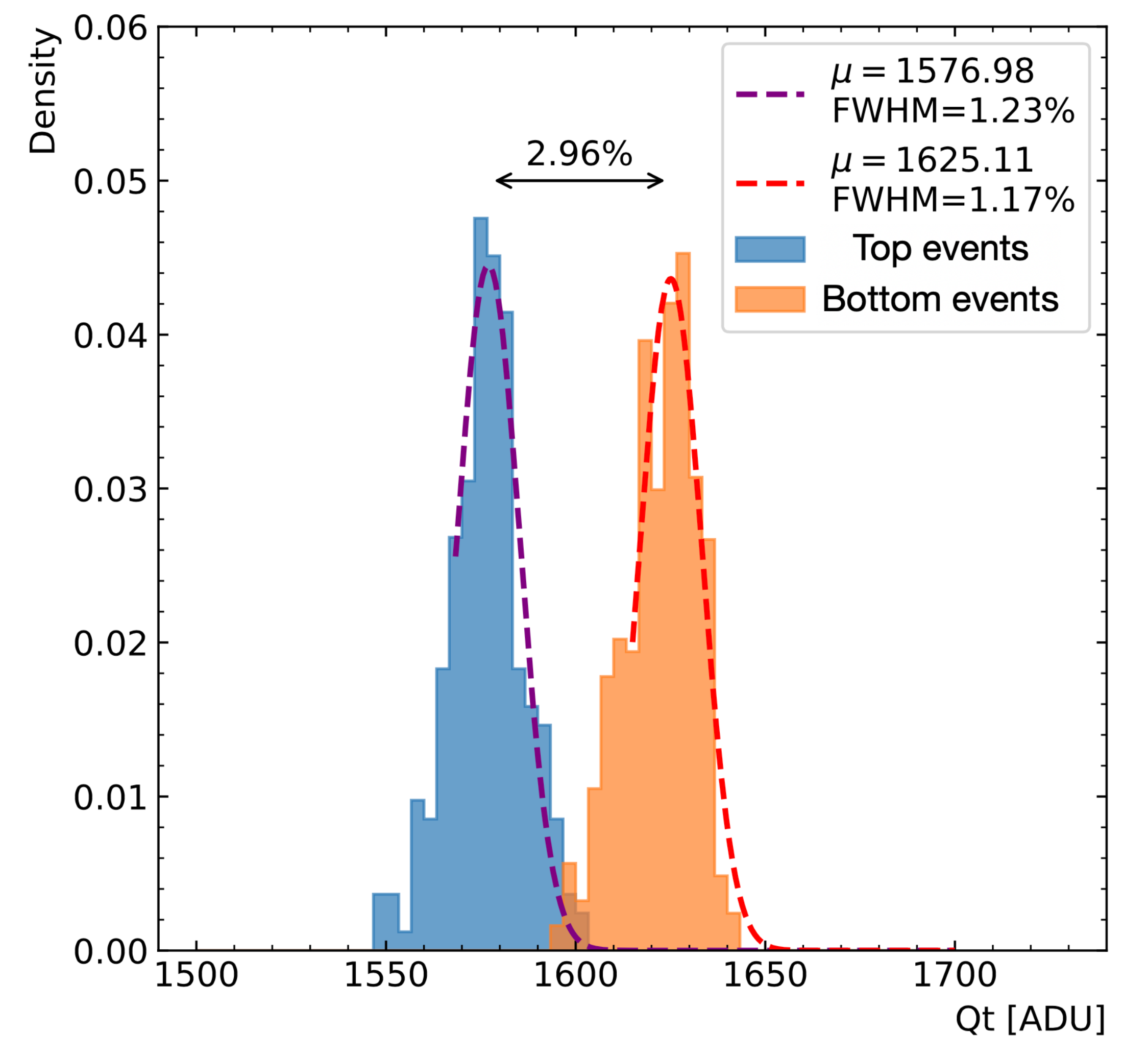}
    \caption{{\it Energy resolution for events contained in the top and bottom hemispheres respectively. 
    \label{fig:correction_voltage:b}}}
\end{figure}

To test the resolution in different regions of 
the detector, a simulation has been performed in which 
5.3~MeV $\alpha$ particles have been uniformly generated throughout 
the detector volume.
The
simulation was performed with a correction voltage of $-150$~V 
applied, and the resolution is shown in Fig.~\ref{fig:correction_voltage:b} for events in which tracks are
entirely contained in either the top (i.e. with the supporting rod) or bottom hemisphere of the detector. A FWHW of $\approx 1.2\%$ is 
observed in both hemispheres, similar to the resolution of 
events from a single source measured in this paper. A $\approx 3\%$
difference is seen in the mean value of $Q_t$ in each hemisphere,
however a further tuning of the correction voltage is expected to fix it it and make the two histograms of Fig.~\ref{fig:correction_voltage:b} overlap.
Given the simulation outcome, and the ongoing R\&D described in
Sec.~\ref{sec:improvements}, it can be reasonably assumed that the desired energy resolution can be achieved throughout the detector.


\section{Future improvements}
\label{sec:improvements}
The search for $\beta\beta0\nu$ decay will feature a xenon-filled SPC  at 40~bar pressure. Based on the experience acquired from this work, a dedicated detector, certified for high-pressure operation, is being designed. 
This detector will be used to study the energy resolution at 
high pressures and, importantly, using xenon gas. 
Given its high cost, the use of xenon will be enabled through the use of a recuperation system. This is currently under preparation at CPPM, and is based on
cryogenic pumping using liquid nitrogen.

A further critical aspect is the xenon gas purity. 
The electron drift velocity in xenon is about two orders of magnitude smaller than in argon, and the probability of a signal reduction due to electronegative impurities is much
larger. 
It is, thus, foreseen that the xenon gas will be flushed through an oxygen capture cartridge to suppress contamination and a recirculation system is under construction to systematically
purify the gas between runs.


The electric field uniformity will be further improved, particularly in the hemisphere containing the rod. 
In addition to the correction electrode approach studied in Sec.~\ref{sec:results_discussion}, further improvements are expected through the application of a degraded high voltage along
the rod, reproducing the voltage distribution of the ideal sphere. This is currently under study by CEA Saclay and Birmingham. Furthermore, the use of a gaseous $^{220}$Rn source would provide uniform distribution of events in the detector volume.



High pressure operation poses challenges in terms of anode voltage and electric field strength at large radii. This problem is resolved by the ``ACHINOS'' multi-anode sensor~\cite{Giganon:2017isb,Giomataris:2020rna}, providing the additional possibility of coarse
track reconstruction. To suppress the effect of avalanche fluctuations to the energy resolution, operation in ionisation mode is also under consideration. These developments would be maximally exploited by corresponding improvements in the read-out electronics, namely an optimized high-voltage filter to further reduce the electronic noise, as discussed in Sec.~\ref{sec:elec}, and the possibility of performing multi-channel readout for the ACHINOS sensor.

\section{Conclusions}
The R2D2 project aims to search for $\beta\beta0\nu$ using a tonne-scale xenon-filled spherical TPC. A crucial element in this endeavour is achieving an energy resolution of approximately 1\% FWHM at 2.458~MeV, the $Q_{\beta\beta}$ of $^{136}$Xe. 
A first prototype was constructed at CENBG
and characterised with argon. 
The main characteristics of the ionisation track, distance and direction relative to the anode, are inferred through pulse shape analysis. 
An energy resolution at the level of 1.1\% was achieved with $\alpha$ particles at 5.3~MeV, corresponding to 1.6\% at 2.458~MeV.
The energy resolution is maintained for both 
point-like and extended energy deposits, a critical finding given that electrons in the $\beta\beta0\nu$ search will 
have lengths of several centimetres. 
A number of potential improvements are identified which, together with the lower W-value of xenon with respect to argon, open the way for the required resolution to be achieved in xenon measurements.

\section*{Acknowledgments}
 The authors would like to thank the IdEx Bordeaux 2019 Emergence program for the OWEN grant for the ``Development of a custom made electronics for a single channel time projection chamber detector aiming at the discovery of neutrinoless double beta decays, and for possible applications in industry''. In addition we thank the CNRS International Emergency Action (IEA) for the ``E-ACHINOS'' grant supporting the collaboration between CENBG and Birmingham University. We thank the CENBG technical staff. We thank M.~Chapellier for providing the Po source and for useful discussions. Discussions with R.~Veenhof are acknowledged. This project has received funding from the European Union's Horizon 2020 research and innovation programme under the Marie Sk\l{}odowska-Curie grant agreement no 841261. K.~Nikolopoulos acknowledges support by the European Research Council (ERC) grant agreement no 714893.

\bibliographystyle{ieeetr}
\bibliography{references}

\end{document}